\documentstyle[12pt]{article}
\def\theequation{\arabic{section}.\arabic{equation}}
\newcommand\T{\theta_{12}}
\newcommand\Tb{{\bar\theta}_{12}}
\newcommand\Z{Z_{12}}
\newcommand\z{z_{12}}
\newcommand\D{{\cal D}}
\newcommand\Db{\overline{\cal D}}
\newcommand\jl{J(Z_1)}
\newcommand\jr{J(Z_2)}
\newcommand\tl{T(Z_1)}
\newcommand\tr{T(Z_2)}

\begin{document}
\renewcommand{\thefootnote}{\fnsymbol{footnote}}
\thispagestyle{empty}
\begin{flushright}
BONN-HE-93-15\\
hep-th/9305070\\
May 1993
\end{flushright}
\begin{center}
{\bf N=2 SUPER - $W_{3}$ ALGEBRA AND N=2 SUPER BOUSSINESQ EQUATIONS}
\vspace{1cm} \\
 E.Ivanov\footnote{E-mail address:
eivanov@pib1.physik.uni-bonn.de},\vspace{0.5cm}\\
{\it Physikalisches Institut, Universitat Bonn} \\
{\it Nussallee 12, D-5300 Bonn 1, Germany}\vspace{0.3cm}\\
and \vspace{0.3cm} \\
{\it Bogoliubov Theoretical Laboratory}\\
{\it JINR -- Joint Institute for Nuclear Research} \\
{\it Dubna, Head Post Office, P.O. Box 79} \\
{\it 101000 Moscow, Russian Federation} \vspace{0.5cm}\\
S.Krivonos\footnote{E-mail address: krivonos@ltp.jinr.dubna.su}
and R.P.Malik\footnote{E-mail address: malik@theor.jinrc.dubna.su}
\vspace{0.5cm}\\
{\it Bogoliubov Theoretical Laboratory}\\
{\it JINR -- Joint Institute for Nuclear Research} \\
{\it Dubna, Head Post Office, P.O. Box 79} \\
{\it 101000 Moscow, Russian Federation} \vspace{1.5cm} \\
{\bf Abstract}
\end{center}

We study classical $N=2$ super-$W_3$ algebra and its interplay with
$N=2$ supersymmetric extensions of the Boussinesq equation in the framework
of the nonlinear realization method and the inverse Higgs - covariant
reduction
approach. These techniques have been previously applied by us in the bosonic
$W_3$ case to give a new geometric interpretation of the Boussinesq
hierarchy. Here we deduce the most general $N=2$ super Boussinesq equation and
two kinds of the modified $N=2$ super Boussinesq equations, as well as the
super Miura maps relating these systems to each other, by applying the
covariant reduction to certain coset manifolds
of linear $N=2$ super-$W_3^{\infty}$ symmetry associated with $N=2$
super-$W_3$.
We discuss the integrability properties of the equations obtained and
their correspondence with the formulation based on the notion
of the second hamiltonian structure.  \vspace{0.5cm}

\begin{center}
{\it Submitted to Int. J. Mod. Phys. A}
\end{center}
\setcounter{page}0
\renewcommand{\thefootnote}{\arabic{footnote}}
\setcounter{footnote}0
\newpage
\section{Introduction}

During the last couple of years, a substantial progress
has been achieved in supersymmetrization of $W$ algebras [1-6]. In
particular,
$N=2$ classical [5] and quantum [6] super-$W_{3}$ algebras
have been constructed. They have attracted a great
deal of interest, mainly in view of their potential applications
in $N=2$ superconformal field theory which is now a subject of
intensive studies (see, e.g., \cite{{G},{KS}}).
The most characteristic feature of the $N=2$ super-$W_3$ algebra is that
it exists for an arbitrary value of the central charge $c$, in contrast to
various minimal $N=1$ extensions of $W_3$ which can be consistently defined
only for specific values of $c$. Thus $N=2$ super-$W_3$ is actually the first
example of a well-defined supersymmetric extension of the nonlinear $W_3$
algebra. The study
of its structure and field-theoretical models associated with it
may shed more light both on the origin of nonlinear (super)algebras
and on the interplay between supersymmetry and $W$ symmetries.

Until now this superalgebra (at the classical level) appeared in the context
of $N=2$ super Toda theories \cite{u7}, super Lax pair formulation \cite{u8}
and
Polyakov ``soldering'' procedure \cite{u5}.
Its realizations on $N=2$ superfields were discussed in \cite{{NY},{IK}}.
In ref. \cite{IK} the first non-trivial hamiltonian flow on
$N=2$ super-$W_3$ ($N=2$ super Boussinesq equation) has been constructed.
Decisive steps towards building a $N=2$ super-$W_3$ string model
have been undertaken in a recent paper \cite{LPWX}.

In the present paper we study $N=2$ super-$W_3$ in the framework of the
nonlinear realizations approach \cite{{NLR},{IH}}, the application of which
to the $W$ type symmetries was initiated in ref.[16-20].

One of the most urgent problems encountered while dealing with
(super) $W$ algebras is to understand in full their geometric origin and,
based on this, to work out convenient general methods for
constructing field-theoretical systems with these algebras
as underlying symmetries. The closely related problem is to
consistently incorporate into a general geometric picture of $W$ algebras
the associated hierarchies of integrable equations (such as the
KdV, Toda, KP and Boussinesq hierarchies and their
superextensions).

There exist several geometric approaches to the $W$ geometry (see, e.g.
\cite{Geom}).
In \cite{{u32},{u32a}} we proposed to treat $W$ symmetries in the
universal geometric language of nonlinear (coset space) realizations
\cite{NLR}.
In this approach a given nonlinear $W_n$ algebra is replaced by some
associate linear infinite-dimensional algebra $W^{\infty}_n$. The latter
is obtained by treating all higher spin composite objects appearing
in the commutators of the basic $W_n$
generators (of spins 2 and 3 in the $W_3$ example) as some independent new
ones. The linearity of $W^{\infty}_n$ symmetry allows to apply to it
the standard techniques of group realizations in homogeneous spaces and
to implement it in a geometric way as a group motion on its
appropriately chosen homogeneous manifolds parametrized by $2D$ space-time
coordinates and infinite sets of
$2D$ fields.
After imposing on these fields the covariant inverse Higgs constraints
\cite{IH} one is left with a finite set of the coset parameter-fields which
define a fully
geodesic $2D$ surface in the original infinite-dimensional coset space. The
standard nonlinear $W_n$ symmetry is recovered as a
particular realization of $W^{\infty}_n$ on this minimal set of
fields.

A remarkable feature of the inverse Higgs effect in the case of
coset space realizations of $W$ symmetries,
besides the fact that it reduces
infinite
towers of the coset parameters to a few essential fields, is that it also
implies some dynamical equations for these fields. This dynamical version
of the inverse Higgs effect was called in \cite{u30} the covariant
reduction. The equations obtained in this way always amount
to the vanishing of some curvature and so are integrable. Moreover,
the well-known Miura maps relating different integrable equations also
turn out to get a nice geometric interpretation as a part of the inverse
Higgs (alias covariant reduction) constraints \cite{u32a}. This provides
a new
systematic way to find
these maps in an explicit form, starting with the defining relations of
given $W$ algebra.

The examples
explicitly elaborated so far are: (1) Liouville equation and its various
superextensions
\cite{{u30},{u31a},{u31}} related to nonlinear realizations of two commuting
light-cone
copies of $W_2$, i.e. Virasoro
symmetry, and superextensions of $W_2$, (2) the $sl_3$ Toda system related
to a nonlinear realization of two light-cone copies of
$W_3$ \cite{u32}, and (3) the Boussinesq equation (together with its modified
versions obtained via Miura-type transformations) related to a nonlinear
realization of one copy of $W_3$ \cite{u32a}. Applications of the coset space
realizations techniques augmented with the inverse Higgs procedure
to the cognate linear $w_{1+\infty}$ symmetry and
its some generalizations were given in \cite{{BS},{SS}} and in
a recent preprint \cite{BIK}.

All this geometric machinery can be rather straightforwardly extended to
$N=2$ super-$W_3$ symmetry, and this is what we do in the present paper.
We show that the covariant reduction approach naturally gives rise to a
$N=2$ superextension of Boussinesq equation and its two modified versions
related to each other via $N=2$ super Miura maps. The equations obtained
amount to the vanishing of some supercurvatures.
In addition to our previous works cited above, this work is
another step in the direction of our main goal of providing a common
geometrical framework for all two-dimensional integrable systems on the
basis of nonlinear realizations of the $W$ type symmetries.

The paper is organized as follows. In
Sec. 2 we recapitulate the essential ingredients of our work \cite{u32a}
on nonlinear realizations of
$W_{3}$ symmetry which will be of need in our subsequent discussion of
$N=2$ super-$W_{3}$ symmetry in a similar context. Then, in Sec. 3, we
recall the basic facts about $N=2$ super-$W_3$ algebra in the formulations
via supercurrents and component currents.
In Sec. 4 we pass to the linear $N=2$ super-$W_3^{\infty}$ symmetry
and construct its coset space realizations generalizing those of
$W_3^{\infty}$. In Sec. 5 we effect the covariant reduction of these coset
spaces and deduce the $N=2$ super Boussinesq equations and the related
super Miura maps as the most essential conditions of the reduction. Sec. 6
is devoted to the comparison with the hamiltonian approach \cite{IK} and
discussion of the integrability properties of the systems obtained. Sec. 7
contains concluding remarks. In Appendices A - C we quote
the basic SOPE's and OPE's of $N=2$ super-$W_3$ algebra,
the (anti)commutation
relations of $N=2$ super-$W_3^{\infty}$ which are used while constructing
the coset space realizations of this symmetry in Sec. 4, and some
unwieldy formulas required in the process of deducing $N=2$ super Boussinesq
equations in Sec. 5.

\section{Preliminaries: nonlinear realization of $W_{3}$ symmetry}

In this Section we sketch the key points
of the coset space realizations of $W_{3}$ algebra.

As $W_3$ algebra is nonlinear, it was unclear how to
generalize to the $W_3$ case the standard techniques of group realizations
in homogeneous spaces \cite{NLR}.
The basic trick invoked in refs.\cite{{u32},{u32a}} to overcome this
difficulty
is to pass to a linear
infinite-dimensional algebra $W_{3}^{\infty}$ in which all the composite
higher spin generators appearing in the commutators of $W_{3}$ are treated
as independent
generators. For instance, the spin 4 composite generator
${\cal J}^{(4)}_m=-\frac{8}{c}\sum_n {\cal L}_{m-n}{\cal L}_n$ appearing
in the classical (centrally extended) $W_{3}$ algebra \cite{SCW}:
\begin{eqnarray}
\left[ {\cal L}_n, {\cal L}_m \right] & = & (n-m){\cal L}_{n+m} +
\frac{c}{12}(n^3-n)\delta_{n+m,0} \nonumber \\
\left[ {\cal L}_n, {\cal W}_m \right] & = & (2n-m){\cal W}_{n+m} \label{21} \\
\left[ {\cal W}_n, {\cal W}_m \right] & = & 16(n-m){\cal J}^{(4)}_{n+m}-
   \frac{8}{3}(n-m)\left[ n^2 + m^2-\frac{1}{2}nm-
4 \right]{\cal L}_{n+m}  \nonumber \\
        & & -\frac{c}{9}(n^2-4)(n^2-1)n \; \delta_{n+m,0} \quad , \nonumber
\end{eqnarray}
together with other higher spin composites
${\cal J}^{(s)}_{n} \;( s=5,6,7,\ldots )$
extend $W_3$ to the $W_{3}^\infty$ algebra as is given below:
\begin{equation} \label{22}
W_{3}^{\infty} = \{ {\cal L}_{n},\; {\cal W}_{n}, \; {\cal J}^{(4)}_{n},
\;...{\cal J}^{(s)}_{n},\;
\ldots \}\;.
\end{equation}
One of the most important subalgebras of (2.2) which plays a crucial role
in the construction of ref. \cite{{u32},{u32a}} contains all the spin
$s$ $(s> 2)$
generators with indices ranging from
$-(s-1)$ to $+ \infty$. It is distinguished in that the explicit central
charge terms drop out from its commutators (though implicit traces of $c$
still remain: e.g., the term linear in ${\cal L}_n$ in the r.h.s. of the
commutator $[ {\cal W}_n,\;{\cal W}_m ]$ is due to $c\neq 0$, one should
make use of
the whole algebra (2.1) while evaluating the commutators with composite
generators, etc). When the above $W_3$ (or $W_3^{\infty}$)
is realized
as a classical symmetry of some $2D$ field-theoretical model
(e.g. the $sl_3$ Toda model),
this subalgebra consists of the infinitesimal group variations with the
parameter-functions regular at $x=0$. In what follows we will deal just
with this truncated version of $W_3^{\infty}$, the ``contact'' $W_3^{\infty}$
(and its $N=2$ superextension). The generators with indices ranging from
$-(s-1)$ to $(s-1)$ constitute a wedge subalgebra $W_\wedge$
in $W_3^{\infty}$. All the composite generators in $W_\wedge$
form an ideal, so that the quotient of $W_\wedge$ by this
ideal is the algebra $sl(3,R)$ \cite{{u32},{u32a}}
\begin{equation} \label{sl}
sl(3,R) \sim  W_{\wedge}/ \{{\cal J}^{(4)}_n, .... {\cal J}^{(s)}_m, ....
\} =
\{ {\cal L}_0, {\cal L}_{\pm 1}, {\cal W}_{0}, {\cal W}_{\pm 1},
{\cal W}_{\pm 2} \}\;.
\end{equation}
Let us point out that the commutation relations of $W_3^{\infty}$ can be
completely restored from the basic $W_3$ relations (2.1). Once this is done,
one may forget that the higher spin generators of $W_3^{\infty}$ were
initially composite and define $W_3^{\infty}$ by its commutation relations.
In practice, it turns out necessary to know only the commutators
involving a few first higher spin generators.

Since $W_3^{\infty}$ symmetry is linear, one may construct the relevant
coset manifolds and define the left action of $W_3^{\infty}$ on them
following general scheme of nonlinear (coset space) realizations.
Thus, by nonlinear realizations of $W_3$
(and $W_n$) symmetry we always mean those of the associate
$W_3^{\infty}$ ($W_n^{\infty}$) symmetry. In the same sense we will
understand nonlinear realizations of $N=2$ super-$W_3$ symmetry.

The specificity of the case at hand is that the relevant coset
manifolds are infinite-dimensional, they are parametrized by the
$2D$ space-time coordinates and an infinite number of $2D$ fields.
However, by imposing an infinite number of
covariant inverse Higgs type constraints on the relevant Cartan forms,
one can reduce
the infinite set of the initial coset parameter-fields to a finite set
of some basic fields and simultaneously obtain a kind of integrable
equations for the latter.

There exist several coset realizations of $W_3^{\infty}$ which differ in
the choice of the
stability subgroup and also in whether one deals with two light-cone copies
of this symmetry or with one copy. The former possibility \cite{u32}
eventually
yields the Lorentz covariant integrable system: the $sl_3$ Toda theory. The
latter option \cite{u32a} (there are three nonequivalent choices of the
relevant realizations) does not respect $2D$ Lorentz covariance and yields
another type of two-dimensional integrable systems, the Boussinesq equation
and two modified Boussinesq equations. Since this case is directly relevant
to the subject of the present paper, we will dwell on it in more detail.

We will restrict our study here
to the realization with the stability subalgebra
${\cal H}_{(2)}$ \cite{u32a} containing a minimal set of generators of
$W_{3}^{\infty}$:
\begin{equation} \label{23}
{\cal H}_{(2)} = \{ {\cal W}_{-1} + 2{\cal L}_{-1}, \;
{\cal J}^{(4)}_{-3},\;\ldots
{\cal J}^{(4)}_{n},
\ldots \;J^{(s)}_{-s+1},\;
\ldots \}\;.
\end{equation}
Two other realizations constructed in \cite{u32a} can be obtained from
this one by putting equal to zero some of the involved coset
parameters or, equivalently, by placing into the stability subalgebra some
of the coset generators, namely, ${\cal L}_0,\;{\cal W}_0$
and ${\cal L}_0, \;{\cal W}_0, \;{\cal L}_1,\;{\cal W}_1, \;{\cal W}_2$,
respectively (following \cite{u32a}, we denote these
subalgebras as ${\cal H}_{(1)}$ and ${\cal H}$). In all cases, the higher spin
generators completing $W_3$ to $W_3^{\infty}$ are placed in the stability
subalgebra. Note that they form a subalgebra on their own (it is an ideal
in ${\cal H}_{(2)} \subset {\cal H}_{(1)} \subset {\cal H}$).

The coset space element $g$ corresponding to the choice (\ref{23})
is parametrized in terms of
coordinates $x,t$ and an infinite tower of the parameter-fields
$(u_0,v_0,u_1,v_1,u_2,v_2,
u,v,\psi_n,\xi_m, n\geq 3, m\geq 4)$ as follows:
\begin{equation}\label{24}
g=e^{t{\cal W}_{-2}}e^{x{\cal L}_{-1}}e^{u{\cal L}_2}e^{v{\cal W}_3}
e^{\sum_{n\geq 3}\psi_n{\cal L}_n}e^{\sum_{n\geq 4}\xi_n {\cal W}_n}
e^{u_1 {\cal L}_1}e^{v_1 {\cal W}_1}e^{v_2 {\cal W}_2}
e^{u_0 {\cal L}_0}e^{v_0{\cal W}_0}\quad .
\end{equation}
Here the "time" coordinate $t$ is linked with the generator
${\cal W}_{-2}$ (which so has a meaning of the time translation generator),
while the spatial coordinate $x$ is associated with the
generator ${\cal L}_{-1}$. All parameter-fields are assumed to be
arbitrary functions of $x$ and $t$, so at this step we are actually dealing
with a two-dimensional
surface embedded in the above coset space, with the parameter-fields as the
embedding functions. The $t$ and $x$ directions on the coset manifold are
entirely
independent of each other because ${\cal W}_{-2}$ commutes with ${\cal
L}_{-1}$.
The $W_3^{\infty}$ symmetry is realized as left shifts of the coset element
(\ref{24}).

The fundamental geometric quantity in the coset space approach
is the Cartan one-form $\Omega=g^{-1}dg$ which can be expressed explicitly
as a sum over the spin $s$ $(s\geq 2)$ generators with indices ranging
{} from $-(s-1)$ to $+\infty$. Then one accomplishes the covariant reduction,
which means that the Cartan
form is restricted to some subalgebra $\widetilde{{\cal H}}_{(2)}$
containing the stability
subalgebra (\ref{23}):
\begin{eqnarray} \label{BCRA2}
\Omega &\Rightarrow & \Omega^{red} \in \widetilde{{\cal H}}_{(2)}\;,
\nonumber \\
\widetilde{{\cal H}}_{(2)} &=& \{ {\cal H}_{(2)}, {\cal W}_{-2},
{\cal L}_{-1} \}\;.
\end{eqnarray}
This procedure is manifestly covariant with respect
to the left action of $W_3^{\infty}$. Putting equal
to zero
all the components of the Cartan form which do not belong to the covariant
reduction subalgebra $\widetilde{{\cal H}}_{(2)}$
leads to expressing all higher spin
parameter-fields in terms of the two essential fields $u_{0}$
and $v_{0}$:
$$
u_1=\frac{u'_0}{2} \quad , \quad v_1=\frac{v'_0}{3} \quad , \quad
v_2=\frac{1}{12}\left( v_0''+u_0'v_0'\right) \quad , \quad
u=\frac{1}{6}\left[ u_0''+\frac{1}{2}\left( u_0'\right)^2
+\frac{8}{3}\left( v_0'\right)^2 \right]
$$
\begin{equation}\label{25}
v=\frac{1}{5}\left[ \frac{1}{12}v_0'''+\frac{1}{12} u_0''v_0'
+\frac{1}{4}u_0'v_0'' +\frac{1}{6}\left( u_0'\right)^2 v_0'
 -\frac{8}{27}\left( v_0'\right)^3  \right]  \quad  etc \quad ,
\end{equation}
where prime stands for $x$ derivative. Simultaneously, for
the essential fields there arise the following dynamical equations:
\begin{equation}\label{26}
\dot{u}_0=-\frac{16}{3}\left[v_0''+2u_0' v_0'\right] \quad , \quad
\dot{v}_0=u_0''-\left( u_0' \right)^2 +\frac{16}{3}\left( v_0'\right)^2\;,
\end{equation}
where dot denotes $t$ derivative.

It is obvious from
equations (\ref{24}) and (\ref{25}) that, whereas the original
coset manifold is
pa\-ra\-me\-tr\-i\-zed
by an infinite number of fields which carry no dynamics, the reduced
manifold is characterized by
only two essential fields subjected to eq. (\ref{26}). Geometrically, this
means that the two-dimensional surface parametrized by $x,t$ and embedded
in the original coset manifold is required to be a {\it geodesic} surface
\cite{u32a}. Most essential points are, first, that this surface
is singled out in a way manifestly covariant under $W_3^{\infty}$ so that
the set $(x,t,u_0 (x,t), v_0(x,t))$ turns out to be closed under the action
of this symmetry, and, second, that
among the conditions defining the surface one finds the set of dynamical
equations (\ref{26}).
Moreover, the latter automatically proves to be equivalent to a
zero-curvature condition for the reduced Cartan form $\Omega^{red}$
as a consequence of the
original kinematical Maurer-Cartan equations for $\Omega$ and the dynamical
inverse Higgs - covariant reduction constraints. Indeed, the reduced Cartan
form, modulo an infinite-dimensional ideal formed by all
higher spin generators of the stability subalgebra, can be
easily found to be
\begin{eqnarray}
\Omega^{red} & = & e^{-2u_0}dt \; {\cal W}_{-2}+
e^{-u_0}\left[ \left( dx+\frac{16}{3}v_0'dt \right) cosh(4v_0 )+
\left( 4u_0'dt \right) sinh(4v_0 ) \right] {\cal L}_{-1}-\nonumber \\
 & & {1\over 2} e^{-u_0} \left[ \left( dx+\frac{16}{3}v_0'dt \right)
sinh(4v_0 )+
\left( 4u_0'dt \right) cosh(4v_0 ) \right] {\cal W}_{-1} \;. \label{27}
\end{eqnarray}
Taking into account that the generators $W_{-2}, L_{-1}$ and $W_{-1}$ form
a three-dimensional subalgebra of the $sl(3, R)$ (\ref{sl}) (modulo an
infinite-dimensional ideal just mentioned), it is
straightforward to check that the Maurer-Cartan equation
$$
d^{ext}\Omega^{red}=\Omega^{red}\wedge\Omega^{red}
$$
leads just to the equations (\ref{26}).

To clarify the meaning of eqs. (\ref{26}), let us see which equations
they entail for the pairs of the composite coset fields $u_1, v_1$ and
$u, v$ via the relations (\ref{25}). It turns out that in both cases we are
left with the closed sets of integrable equations. The first pair of
equations trivially
follows by taking $x$ derivative of (\ref{26}). The second pair reads
\begin{equation}  \label{26a}
\dot{u} = - {160\over 3} v'\;,\;\;\;\; \dot{v} = {1\over 10}u''' -
{24\over 5}u'u\;,
\end{equation}
and it is easily recognized (after proper rescalings)
as the Boussinesq equation. As was explained in
ref. \cite{u32a}, the coset fields $u, v$ have the correct conformal
properties to be identified with the spin 2 conformal stress-tensor and a
primary spin 3 current, respectively. Then relations (\ref{25}) are the
appropriate Miura maps projecting $u,\; v$ onto the sets of the spin 1
currents and the spin 0 scalar fields. With this in mind,
eqs. (\ref{26})
(as well as the corresponding equations for $u_1, v_1$) can
be called modified Boussinesq equations. Note that eqs. (\ref{26a}) and the
equations for $u_1, v_1$ can be independently obtained by applying the
covariant reduction techniques to two other coset manifolds of
$W_3^{\infty}$ mentioned above, with zero-curvature representations on
$sl(3,R)$ (\ref{sl}) and a five-dimensional Borel subalgebra of the latter
(for details see ref.\cite{u32a}). We also
point out that all these equations are covariant by construction under
the $W_3^{\infty}$ symmetry. As was shown in \cite{u32a},
while applied to the essential coset fields,
the $W_3^{\infty}$ transformations coincide with those of $W_3$ symmetry
which thus proves to be a particular realization of $W_3^{\infty}$.

To summarize, in the coset space approach the Boussinesq and
modified Boussinesq equations as well as the corresponding Miura maps
naturally arise within the single geometric procedure, the
covariant reduction on homogeneous spaces of infinite-dimensional
linear symmetry $W_3^{\infty}$ associated in a definite way
to $W_3$ algebra. Taking as an input the defining relations of $W_3$
algebra and further employing a number of geometrically motivated
prescriptions, we obtain as an output the above equations
together with the zero-curvature representation for them and
the explicit form of Miura maps. Moreover, the covariance of these
equations with respect to $W_3$ symmetry becomes evident and one
may explicitly find the $W_3$ transformations of the involved fields.
The basic spin 2 and spin 3 $W_3$ currents as well as the related to
them via Miura maps spin 1 and spin 0 fields get a novel geometric
interpretation as coordinates of the $W_3^{\infty}$ coset manifolds.

In the next sections we will discuss how all this can be generalized to
the case of $N=2$ super-$W_3$ symmetry. We will derive $N=2$ superextensions
of Boussinesq and modified Boussinesq equations and the relevant $N=2$
superfield Miura maps.

\setcounter{equation}0

\section{$N=2$ super-$W_{3}$ algebra}

In this section, we briefly recapitulate salient features
of the classical
$N=2$ super-$W_{3}$ algebra \cite{u5} required for the
realization of this algebra through the coset superspace construction.

All the basic currents of $N=2$ super-$W_{3}$ algebra are accomodated by
the spin 1 supercurrent $J ( Z )$
and the spin 2 supercurrent $T( Z )$, where
$Z \equiv (x, \theta, \bar{\theta})$ are coordinates of $N=2,\;1D$
superspace. Indeed, the components of these $N=2$ superfields
carry, respectively, the conformal spins $(1,3/2,3/2,2)$ and
$(2,5/2,5/2,3)$, precisely as the currents generating $N=2$ super-$W_3$.
The supercurrent $J$ generates $N=2$ super Virasoro algebra, while
$T$ can be chosen to be primary with respect to the latter \cite{IK}.
The closed set of SOPE's between these supercurrents has been explicitly
written in \cite{IK}. We quote them in Appendix A.

The component currents appearing in the $\theta, \bar{\theta}$
decomposition of $T$ are related to their
counterparts from ref.\cite{u5} via a nonlinear
redefinition,
because the second set of currents is assembled into a non-primary
$N=2$ supermultiplet.
Explicitly, the relation between the currents present in $J$ and $T$ and
the currents of ref. \cite{u5} is as follows
\begin{eqnarray}
 J| \equiv J = 4{\cal J} & , & T| \equiv \widetilde{T} =
   {\cal T}+4\widetilde{\cal T}-\frac{128}{c}{\cal J}^2 \nonumber \\
 {\cal D}J| \equiv \bar{G} = \bar{\cal G} & , & {\cal D}T| \equiv \bar{U}
     =\frac{3}{4}\bar{\cal U} - \frac{64}{c}{\cal J}\bar{\cal G} \nonumber \\
- \bar{\cal D} J| \equiv G ={\cal G} & , & -\bar{\cal D} T| \equiv U =
\frac{3}{4}{\cal U} - \frac{64}{c}{\cal J} {\cal G} \label{31} \\
 \frac{1}{2}\left[ \bar{\cal D},{\cal D} \right] J| \equiv T =
    {\cal T}+\widetilde{\cal T} & , &
\frac{1}{2}\left[ \bar{\cal D},{\cal D} \right] T| \equiv W =
  \frac{3}{4}{\cal W}+\frac{32}{c}\left( {\cal T}+
  4\widetilde{\cal T}-\frac{128}{c}{\cal J}^2
 \right) {\cal J} +\frac{40}{c}{\cal G}\bar{\cal G} \; ,  \nonumber
\end{eqnarray}
where $|$ means restriction to the $\theta, \bar{\theta}$ independent parts
and the currents written in calligraphic obey OPE's of ref. \cite{u5}. The
covariant spinor derivatives are defined by
\begin{eqnarray} \label{DefD}
{\cal D}_{\theta} = \frac{\partial}{\partial \theta} -{1\over 2}
\bar{\theta}\partial_x\;,\;\;\;\;&& \bar{{\cal D}}_{\theta} =
\frac{\partial}{\partial \bar{\theta}} -{1\over 2} \theta \partial_x
\nonumber \\
\{ {\cal D}, \bar{{\cal D}} \}=-\partial_x\;,
&&{{\cal D}}^2={\bar {{\cal D}}}^2=0\;.
\end{eqnarray}

The explicit form of OPE's between the currents contained in
$J$ and $T$ is given in Appendix A. It seems useful to make here a few
comments on them and their relation to those given in ref. \cite{u5}.

The property that the supercurrent $T$ is primary with respect to the
$N=2$ conformal supercurrent manifests itself as well on the level of
OPE's. For instance, the following OPE's (the currents in the r.h.s.
are always evaluated at the second argument):
\begin{eqnarray}
T(z_1)\widetilde{T}(z_2) & = &  {2\widetilde{T} \over (z_1-z_1)^2} +
        {\widetilde{T}' \over {z_1-z_2}} \nonumber \\
J(z_1)\widetilde{T}(z_2) & = & 0 \label{32}
\end{eqnarray}
demonstrate that $\widetilde{T}$ is a spin 2 primary field with
respect to $T$ and it carries
zero $U(1)$ charge.

One more remark concerns OPE's of fermionic currents. OPE's
$\sim G(z_1) G(z_2),\;\bar{G}(z_1) \bar{G}(z_2)$ are vanishing but this
is not
the case for analogous OPE's involving the currents $U,\;\bar{U}$.
These are as follows:
\begin{equation}\label{33}
U(z_1) U(z_2)  =  {1\over c} \left( {{80GU-32 G G'} \over {z_1-z_2}} \right)
\quad , \quad
\bar{U}(z_1)\bar{U}(z_2)  =
 - {1\over c}\left( {{80\bar{G}\bar{U}+32\bar{G} \bar{G}'} \over {z_1-z_2}}
\right) \;.
\end{equation}
This fact is essential for validity of the following
graded Jacobi identity
\begin{equation} \label{34}
\left[ U_p, \left\{ {\bar G}_r, U_s \right\} \right]+
\mbox{graded cyclic} = 0
\end{equation}
(for definition of the $N=2$ super-$W_3$ generators see Appendix B).
Note that among the OPE's explicitly written down in
\cite{u5} analogous important OPE's for ${\cal U},\; \bar{{\cal U}}$
were missing. Using the relations (3.1), these OPE's can be found to be
\begin{equation}\label{35}
{\cal U}(z_1) {\cal U}(z_2)  = {64\over c} \left(
{{ {\cal G}{\cal U}-\frac{8}{3}{\cal G} {\cal G}'} \over {z_1-z_2}} \right)
\quad
\mbox{ and } \quad
\bar{\cal U}(z_1)\bar{\cal U}(z_2)  =
  -{64\over c}\left( {{ \bar{\cal G}\bar{\cal U}+\frac{8}{3}\bar{\cal G}
\bar{\cal G}'}
\over {z_1-z_2}} \right) \;.
\end{equation}

Proceeding from the OPE's listed in Appendix A, it is
straightforward to compute the
(anti) com\-m\-u\-tators of the
$N=2$ super-$W_3$ generators defined as Laurent modes of the
component currents. These relations are quoted in Appendix B.
They are of primary use while treating $N=2$ super-$W_{3}$ algebra in the
framework of the coset space realizations method. This will be the subject
of the next Section.

\setcounter{equation}0

\section{Nonlinear realizations of $N=2$ super-$W_{3}$}

In this Section, following the prescriptions outlined
in ref. \cite{{u32},{u32a}} and Sec. 2, we define a linear
infinite-dimensional $N=2$ super-$W_3^{\infty}$ algebra from
$N=2$ super-$W_{3}$ and
discuss the choice of the relevant stability subalgebras
and construction of the associate
infinite-dimensional coset supermanifolds.

\subsection{From $N=2$ super-$W_{3}$ to $N=2$ super-$W_{3}^{\infty}$}

Like in the $W_3$ case, in order to construct
coset space realizations of $N=2$ super-$W_{3}$ symmetry one
should firstly define a linear infinite-dimensional
$N=2$ super-$W_{3}^{\infty}$ algebra.
Such a superalgebra (hereafter denoted as $sW_3^{\infty}$)
can be obtained by treating as independent all the higher spin
composite generators appearing in the (anti)commutators of
the basic $N=2$ super-$W_{3}$ generators. Applying to Appendix B,
we find that $sW_3^{\infty}$ is constituted by the
following generators:
\begin{equation}\label{41}
sW_3^{\infty}=\{\; J_n,G_r,\bar{G}_r,L_n,\widetilde{L}_n,U_r,\bar{U}_r,
W_n,B^{(2)}_n,V_r^{(5/2)},\bar{V}_r^{(5/2)},\ldots,
\Phi^{(s)}_n,\Xi_r^{(s)},\bar{\Xi}_r^{(s)},\ldots,  \}\;.
\end{equation}
Here the generators
$ J,G,\bar{G},L,\widetilde{L},U,\bar{U} \mbox{ and } W $
are the basic generators of $N=2$ super-$W_{3}$ (they come from the
$\theta$ and $x$ decompositions of the supercurrents $J, T$) and
$B^{(2)}_n,V_r^{(5/2)},\bar{V}_r^{(5/2)},\ldots $
are the generators coming from composite currents of conformal spins
$(2,7/2,7/2,3)$ the explicit form of which is also given in
Appendix B. The letters
$(\Phi^{(s)}_n,\Xi_r^{(s)},\bar{\Xi}_r^{(s)},\ldots )$ stand
for still higher spin composite generators of $sW_3^{\infty}$.

Just as in the bosonic case, we will deal not with the whole $sW_3^{\infty}$
but with its ``contact'' subalgebra which is singled out by restricting
the indices of all spin $s\; (s\geq 1)$
generators of $sW_3^{\infty}$ to vary from $-(s-1)$ to $+\infty$. Like its
bosonic prototype, the contact $sW_3^{\infty}$ contains no explicit central
charge terms. In what follows we will, as a rule, omit the adjective
``contact''.

The reflection symmetry, $n\rightarrow (-n)$
and $r\rightarrow (-r)$, inherent in the full superalgebra
$sW_3^{\infty}$ guarantees the existence
of an infinite-dimensional
wedge subalgebra $sW_{\wedge}$ in the contact $sW_3^{\infty}$. It
encompasses all the spin $s$ generators
$(s\geq 1)$ with indices varying from $-(s-1)$ to $ (s-1)$.

Inspection of the structure of $sW_{\wedge}$ shows that it contains
an infinite-dimensional ideal collecting all composite generators.
For our further purposes it will be essential that the
quotient of $sW_{\wedge}$ over this ideal is the superalgebra
$sl(3|2)$:
\begin{eqnarray} \label{osp}
sl(3|2) \sim  sW_{\wedge} / \{ B^{(2)}_n,V_r^{(5/2)},\bar{V}_r^{(5/2)},
\ldots \}
&=&
\{ J_0, L_0, L_{\pm 1}, \widetilde{L}_0, \widetilde{L}_{\pm 1},
W_{0}, W_{\pm 1}, W_{\pm 2},  \nonumber \\
&& G_{\pm 1/2}, \bar{G}_{\pm 1/2},
 U_{\pm 1/2}, U_{\pm 3/2},
\bar{U}_{\pm 1/2}, \bar{U}_{\pm 3/2} \}\;.
\end{eqnarray}
It is an obvious generalization of the quotient algebra $sl(3,R)$ (\ref{sl}).

It is worthwhile to mention here that, in general, generators associated
with the currents of spins $2,\;3$ and $5/2$ can be modified
by adding generators coming from
the same spin composite currents, such as $J^2,\;
TJ,\; \widetilde{T}J, \; J^3,\; G\bar{G}\;, JG $, and $J\bar{G}$,
without changing the structure of the quotient algebra (4.2).
Usefulness of this
statement will become more lucid and transparent in Sec.5 in the process
of construction of $N=2$ super Boussinesq equations (see also Subsec.4.3).

\subsection{Stability subalgebras}

In accordance with the discussion in Sec.2, the important step in
defining a coset space realization of the $sW_3^{\infty}$ symmetry
is to choose an appropriate stability subgroup. It is beyond
the scope of the present paper to list all possible candidates for this
role. Like in the $W_3$ case, we require that all the composite
generators are in the stability subalgebra. Then there remain a few
possibilities which are easy to analyze.

The maximally
enlarged subalgebra of this sort is obtained by putting together the
composite generators and the $sl(3|2)$ generators (\ref{osp}) from
$sW_{\wedge}$
\begin{equation} \label{CRA0}
s\widetilde{{\cal H}} = \{\; sl(3|2) \oplus \mbox{Higher spin generators}
\}\;.
\end{equation}
One may check that the higher spin generators still form an ideal
in (\ref{CRA0}), so the $sl(3|2)$ can equivalently be regarded as a
quotient of (\ref{CRA0}) by this ideal
\begin{equation} \label{sl32}
sl(3|2) \sim  s\widetilde{{\cal H}}/
\{  B^{(2)}_n,V_r^{(5/2)},\bar{V}_r^{(5/2)},\ldots \} \; .
\end{equation}
However, (\ref{CRA0}) is not
quite appropriate candidate for the stability subalgebra (like its
$W_3$ prototype $sl(3,R)$, \cite{u32a}) as it contains the translations
generators $L_{-1}$ and $W_{-2}$ which should certainly be in the coset.
Actually, in order to have a manifest $N=2$ supersymmetry we are led
to place in the coset the generators
$W_{-2}, L_{-1}, G_{-1/2}, \bar{G}_{-1/2}$ with which the coordinates
of $N=2$, $2D$ superspace $(t, x, \theta, \bar{\theta}) \equiv (t, Z)$
will be associated
as the relevant coset parameters (see Subsec.4.3)\footnote{$s\widetilde
{{\cal H}}$ can
still be chosen as a covariant reduction subalgebra, see Sec. 5.}.
After extracting these
generators from (\ref{CRA0}) we are left with a set which is not closed;
to get a closed set we need to transfer into the coset more $sl(3|2)$
generators. We wish all the $sl(3|2)$ generators with negative
indices except those mentioned above to lie in the stability subalgebra,
because otherwise we would need to extend the $N=2$ superspace by some new
coordinates associated with these generators. The maximally possible closed
set which meets
this criterion is as follows
\begin{eqnarray} \label{StA0}
s{\cal H} &=& \{ W_{-1} +{3\over 2} L_{-1}, L_{-1} - \widetilde{L}_{-1},
U_{-3/2}, \bar{U}_{-3/2}, G_{-1/2} + U_{-1/2}, \bar{G}_{-1/2} -
\bar{U}_{-1/2},
J_0, L_0, \widetilde{L}_0, W_0, \nonumber \\
       && G_{1/2} + {1\over 2} U_{1/2}, \bar{G}_{1/2} - {1\over 2}
\bar{U}_{1/2}, L_1 - {1\over 4}\widetilde{L}_1, W_1, W_2,
\mbox{Higher spin generators} \}\;.
\end{eqnarray}
This superalgebra is an analog of the stability algebra ${\cal H}$ of
the $W_3$ case \cite{u32a} and contains ${\cal H}$ as a bosonic subalgebra.
However, the analogy is not literal because ${\cal H}$ is less than
$\widetilde{{\cal H}} = \{ sl(3,R) \oplus \mbox{Higher spin $W_3^{\infty}$
generators} \}$ just by the $t$ and $x$ translations
generators $W_{-2}$ and $L_{-1}$ while the quotient of
$s\widetilde{{\cal H}}$
by $s{\cal H}$ includes additional $sl(3|2)$ generators besides those of
the $t,x$ translations and $N=2$ supertranslations. Actually, this subtlety
is directly related to the structure of the superalgebra $sl(3|2)$ and we
will make some further comments on it in Sec. 5.

By comparing the contents of $s{\cal H}$ and $sW_3^{\infty}$ it is
evident that the coset superspace associated with this choice is
infinite-dimensional similarly to the bosonic case.
Since all the higher spin generators have been placed in the stability
subgroup, in the coset there remain only the proper sets of
generators related to the basic currents of $N=2$ super-$W_3$, i.e. those
with spins $(1, 3/2, 3/2, 2)$ and $(2, 5/2, 5/2, 3)$.

Two other relevant stability subalgebras
can be obtained as proper
truncations of (\ref{StA0}). The related coset manifolds are also
infinite-dimensional. An analog of the stability subalgebra
${\cal H}_1$ of the $W_3$ case is $s{\cal H}_{(1)}$ defined as follows
\begin{eqnarray} \label{StA1}
s{\cal H}_{(1)}& =& \{ W_{-1} +{3\over 2} L_{-1}, L_{-1} - \widetilde{L}_{-1},
U_{-3/2}, \bar{U}_{-3/2}, G_{-1/2} + U_{-1/2},
  \bar{G}_{-1/2} - \bar{U}_{-1/2}, \nonumber \\
    & & J_0, L_0, \widetilde{L}_0, W_0, \mbox{Higher spin generators} \}\;.
\end{eqnarray}
It should be noticed that, similar to the bosonic case \cite{u32a},
just the specific combinations of
generators indicated in eq.(\ref{StA1}) form the subalgebra.
Furthermore, it is the maximally
possible stability subalgebra which still forms a closed set with the
$N=2$
superspace translations generators
$W_{-2}, L_{-1}, G_{-1/2}, \bar{G}_{-1/2}$. Precisely this set of
generators is the covariant reduction subalgebra appropriate for the
geometric derivation of $N=2$ super Boussinesq and modified
super Boussinesq equations, as well as Miura maps of the $N=2$
$W_3$ supercurrents onto the spin $1/2$ supercurrents \cite{IK}
(we denote this subalgebra with $s\widetilde {{\cal H}}_{(1)}$).
To deduce further Miura maps onto scalar superfields \cite{{NY},{IK}} in the
coset space approach,
one needs to pass to the realization with a smaller stability
subalgebra, namely,
\begin{eqnarray} \label{StA2}
s{\cal H}_{(2)}& =& \{\; U_{-3/2},\;
\bar{U}_{-3/2},\;
G_{-1/2}+U_{-1/2},\;  \bar{G}_{-1/2}-\bar{U}_{-1/2},\; \nonumber \\
 & &  L_{-1}-\widetilde{L}_{-1},\;
 W_{-1}+\frac{3}{2}L_{-1},\;\; \mbox{Higher spin generators} \}\;,
\end{eqnarray}
which is obtained from $s{\cal H}_{(1)}$ by removing the generators
$W_0, L_0, \tilde{L}_0, J_0$ (which are thus transferred into the coset).
It is
an analog of ${\cal H}_2$ of the $W_3$ case (eq. (\ref{23})). This
set and the aforementioned $N=2$ superspace translations generators
still form a subalgebra $s\widetilde{{\cal H}}_{(2)}$, the covariant reduction
to which yields the
whole set of $N=2$ super Boussinesq equations and Miura maps.

\subsection{Construction of the coset supermanifolds}

As was mentioned in the above discussion,
for our ultimate purpose
of getting $N=2$ superextensions of the Boussinesq equations (\ref{26}),
(\ref{26a}) and the relevant super Miura maps it is enough to consider
the realization of $sW_3^{\infty}$ which is associated with the
stability subalgebra $s{\cal H}_{(2)}$ defined
in eq. (\ref{StA2}). We will comment
on the realizations corresponding to the stability subalgebras (\ref{StA0})
and (\ref{StA1}) in Sec. 5.

An element of the coset supermanifold of $sW_3^{\infty}$ symmetry
corresponding to the choice of the stability subalgebra (\ref{StA2})
can be parametrized as follows:
\begin{eqnarray} \label{CE}
g_{(2)} & = & e^{tW_{-2}}e^{xL_{-1}}e^{\theta G_{-1/2}+
\bar{\theta}\bar{G}_{-1/2}}
e^{\phi J_1}e^{\mu G_{3/2}+\bar{\mu}\bar{G}_{3/2}}
e^{u_2L_{2}}e^{\tilde{u}_2\tilde{L}_{2}}e^{\phi_2 J_2}
e^{\mu_1 G_{5/2}+\bar{\mu}_1\bar{G}_{5/2}}
e^{\nu_1 U_{5/2}+\bar{\nu}_1\bar{U}_{5/2}}
\ldots \nonumber \\
 & &
e^{\chi G_{1/2}+\bar{\chi}\bar{G}_{1/2}}
e^{uL_{1}}e^{\tilde{u}\tilde{L}_{1}}
e^{v_1W_{1}}
e^{\nu U_{3/2}+\bar{\nu}\bar{U}_{3/2}}
e^{v_2W_{2}}
e^{-\frac{1}{2} ( \xi U_{1/2}+\bar{\xi}\bar{U}_{1/2})}
e^{u_{0}L_0}e^{\tilde{u}_0 \tilde{L}_0} e^{\phi_0 J_0}
e^{v_0 W_0} \;.
\end{eqnarray}
Here the factor $(-1/2)$
before the coset parameters $\xi, \bar{\xi}$ has been introduced for further
convenience. The time coordinate $t$ with dimension $(cm)^2$ and
the spatial coordinate $x$ with dimension $(cm)^1$, like in the bosonic
case, are associated with the
$t$ and $x$ translation generators $W_{-2}$ and $L_{-1}$.
A new point is
the presence of the Grassmann
coordinates $\theta$ and $\bar\theta$ (of dimension $(cm)^{1/2}$)
which are associated with
the fermionic generators $G_{-1/2}$ and $\bar{G}_{-1/2}$ and extend
$(t, x)$ to a $N=2,\; 2D$ superspace
$(t,Z) \equiv (t,x,\theta, \bar{\theta})$
(the anticommutator of
$G_{-1/2}$ and $\bar{G}_{-1/2}$ is just $L_{-1}$, see Appendix B).
An infinite tower of the remaining coset parameters
$(\chi,\bar\chi,\xi,\bar\xi,u,{\tilde u}, \phi,v,u_1,{\tilde  u}_1,\ldots )$,
with spins being various integers and half-integers, are assumed to be
superfields given on this superspace. The group $sW_3^{\infty}$ (to be more
precise, its ``contact'' subgroup, see Subsection 4.1) acts
on the element $g_{(2)}$ as left shifts, which induces an infinite sequence
of
symmetry transformations of the coset parameters. The important point
about these transformations is that in general they mix the $N=2$ superspace
coordinates $t, x, \theta, \bar{\theta}$ with the parameter-superfields,
i.e. these coordinates alone {\it do not form} an invariant subspace of
$sW_3^{\infty}$, quite analogously to the $W_3$ case. In principle,
the $sW_3^{\infty}$ transformations of the coset parameters, at least
those corresponding to
the basic spins generators, can be found explicitly with making use of
the basic (anti)commutation relations given in Appendix B and some additional
relations involving composite generators. In what follows we will not
be interested in the explicit form of these transformations.
For our purposes it will be sufficient to know that at each step we
preserve the $sW_3^{\infty}$ covariance and so the final relations
($N=2$ super Boussinesq equations, the relevant super Miura maps, ...)
certainly respect this symmetry.

Let us offer a few comments concerning the order of the group
factors in (\ref{CE}) and the choice of the $t$ translations generator.

Just because of the special arrangement of the $t$, $x$ and $\theta$
exponentials in the element (\ref{CE}) the remaining coset parameters
behave
as scalars
under $t$ translations, $x$ translations and rigid supertranslations realized
as left shifts of (\ref{CE}). This is the reason why
they can be consistently treated as $N=2$ superfields given on the
superspace $(t, Z)$. Note that
the $t$ translation
generator $W_{-2}$ commutes (like in the bosonic case) with
all other translation operators, namely $L_{-1}$, $G_{-1/2}$ and
$\bar{G}_{-1/2}$. Physically this can be expressed
as the statement that, on the coset supermanifold,
translations along the $t$ direction are entirely
independent of translations along the $x$, $\theta$ and
$\bar\theta$ directions.

It is interesting that, in contrast to the bosonic case,
now there is a freedom in the
definition of the $t$ translations generator. Namely, as such a generator
one may equally choose the following linear combination of $W_{-2}$
and one of the composite generators present in the stability subalgebra:
\begin{equation} \label{Wnew}
W_{-2}^{(\alpha)} = W_{-2}+\alpha\bigl(TJ- G{\bar G}\bigr)_{-2}\;.
\end{equation}
Here $\alpha$ is an arbitrary real parameter.
It can be readily checked that this generator, like $W_{-2}$, commutes
with all the rest of the coordinate translations generators. Note that
the extra term in (\ref{Wnew}) is the unique combination which has the same
conformal dimension as $W_{-2}$ and meets the aforementioned
commutativity requirement. After substituting $W_{-2}^{(\alpha)}$ for
$W_{-2}$ in (\ref{CE}) we obtain a
one-parameter family of the coset space realizations of $sW_3^{\infty}$
with the same stability subgroup.
Note that the quotient algebra $sl(3|2)$ defined in Subsec. 4.1
remains the same even if we replace $W_{-2}$ by $W_{-2}^{(\alpha)}$
in the wedge superalgebra $sW_\wedge$ .

This freedom will be used in Sec. 5 to derive the most general $N=2$
super Boussinesq equation in the framework of the coset space approach.

\subsection{Cartan forms}

As was mentioned in Sec. 2, the fundamental geometrical quantity
which determines curvature, torsion and other characteristics of a
(super)group coset manifold is the differential covariant Cartan one-form
$(\Omega )$.
For simplicity, we specialize here to the case $\alpha = 0$.
For the coset supermanifold in question, the Cartan form is introduced as
follows:
\begin{eqnarray}
\Omega_{(2)} & \equiv & g^{-1}_{(2)}dg_{(2)} = \sum_{n=-2}^{\infty}w_nW_n+
\sum_{n=-1}^{\infty}l_nL_n + \sum_{n=-1}^{\infty}{\tilde l}_n{\tilde L}_n +
\sum_{r=-1/2}^{\infty}g_rG_r + \sum_{r=-1/2}^{\infty}{\bar g}_r{\bar G}_r
 \nonumber \\
& &+\sum_{r=-3/2}^{\infty}f_rU_r + \sum_{r=-3/2}^{\infty}{\bar f}_r
{\bar U}_r
+\sum_{n=0}^{\infty}j_nJ_n+\mbox{ Higher spin contributions}\;, \label{45}
\end{eqnarray}
where we have decomposed $\Omega_{(2)}$ over the $sW_3^{\infty}$
generators. The differentiation in eq.(\ref{45}) is with respect
to the coordinates $t, x, \theta, \bar{\theta}$. One may divide
$\Omega_{(2)}$ further into the coset and
stability subalgebra parts, singling out in the r.h.s. of (\ref{45}) the
appropriate combinations of generators. For the covariance of the
inverse Higgs - covariant reduction procedure it will be essential
that the coefficient superforms associated with coset generators transform
homogeneously. Note that all the forms associated with the higher spin
generators belong to the stability
subalgebra part of $\Omega_{(2)}$ and they will never appear explicitly
in the subsequent consideration.

The evaluation of these forms uses the (anti)commutation relations given
in Appendix B and is straightforward though a bit tiresome.

First few coset forms, up to finite rotations by the group factors
with generators $L_0, \tilde{L}_0, J_0, W_0$ standing on the
right end of $g_{(2)}$ in (4.8),  are as follows:
\begin{eqnarray} \label{FCF}
w_{-2} & \sim & dt  ,   \nonumber \\
l_{-1} & \sim & \Delta x + \left( 12v_1-\frac{3}{2}\xi\bar\xi-
6\bar\chi \xi -6 \chi\bar\xi \right) dt , \\
{\tilde l}_{-1} & \sim & \left( -2 \phi -3 \chi\bar\chi -3 v_1+
 \frac{3}{4} \xi\bar\xi +3 \bar\chi \xi +3 \chi\bar\xi \right) dt
 \mbox{ etc. }\;,\nonumber
\end{eqnarray}
where
\begin{equation}
\Delta x \equiv dx -\frac{\bar\theta d \theta+ \theta d \bar\theta }{2}
\end{equation}
is the covariant (with respect to $N=2$ superconformal transformations)
differential
of $x$. In what follows we will always expand differentials of the
coset parameters-superfields over the covariant set $dt,\;d\theta, \;
d\bar{\theta},\; \Delta x$ according to the rule
\begin{equation} \label{dif}
d = dt\;\partial_t + \Delta x \;\partial_x + d\theta \;{\cal D}_{\theta} +
d\bar{\theta}\; \bar{{\cal D}}_{\theta}\;,
\end{equation}
where the spinor derivatives ${\cal D}, \bar{\cal D}$ have been defined
in eqs. (\ref{DefD}). Note that the arrangement of the exponentials in
eq. (\ref{CE}) is such that
in the Cartan forms there explicitly appear only the covariant
differentials of coordinates and no explicit $t$, $x$ and $\theta$'s can
appear (this is of course a direct consequence of the fact that the coset
parameter-superfields behave as scalars under the $t$ translations and rigid
$N=2,\;1D$ supersymmetry and
everything should clearly be covariant with respect to these symmetries).
The projections of the
whole Cartan form (\ref{45}) and its coefficients on the above set of
differentials are defined by (for the moment we omit the subscript $(2)$ of
$\Omega$ as the subsequent relations are of universal validity)
\begin{equation}  \label{DecO}
\Omega = \Omega_t\;dt + \Omega_x \;\Delta x + \Omega_{\theta} \;d\theta
+ \Omega_{\bar{\theta}}\;d\bar{\theta}\;.
\end{equation}
The Maurer-Cartan equation for the Cartan form $\Omega$,
\begin{equation}  \label{MC}
d^{ext} \Omega = \Omega \wedge \Omega\;,
\end{equation}
implies a number of
useful general identities for these projections, in particular,
\begin{equation} \label{MC1}
\Omega_x = \bar{D} \Omega_{\theta} + D\Omega_{\bar{\theta}} - \{
\Omega_{\theta}, \Omega_{\bar{\theta}} \}
\end{equation}
amounts to the fact that the $\Delta x$ projections
are dependent quantities.

As examples of more complicated (super)forms we quote the
forms which begin with
the differentials of the coset superfields $u_0$, $\tilde{u}_0$, $v_0$ and
$\phi_0$:
\begin{eqnarray}
l_0 & = & du_0 +d\theta \bar\chi +d\bar\theta \chi -2u \Delta x + dt (l_0 )_t
                \nonumber \\
{\tilde l}_0 & = & d{\tilde u}_0 -
      \frac{1}{2}(d\theta -\chi \Delta x )\bar\xi +
      \frac{1}{2}(d\bar\theta -\bar\chi \Delta x )\xi
        -2\tilde u \Delta x + dt ({\tilde l}_0)_t \nonumber \\
\omega_0 & = & d v_0 -3v_1\Delta x +\frac{1}{2}\xi\bar\xi \Delta x -
   \frac{1}{2}(d\bar\theta -\bar\chi \Delta x)\xi -
   \frac{1}{2}(d\theta -\chi \Delta x)\bar\xi +dt (\omega_0 )_t \label{f12}\\
j_0 & = & d\phi_0 - \phi \Delta x-\frac{1}{2}\chi\bar\chi \Delta x
    -\frac{1}{2}\xi\bar\xi \Delta x +
     \frac{1}{2}d\theta\bar\chi -\frac{1}{2}d\bar\theta \chi+
dt(j_0)_t \nonumber
\end{eqnarray}
Here $(l_0)_t,({\tilde l}_0)_t, (\omega_0)_t$ and $(j_0)_t$ are
complicated expressions which collect
many contributions including those from
the generators associated with the composite currents.
These expressions are quoted in Appendix C.
They play an important role in obtaining $N=2$
super-Boussinesq equations for the superfields
$u_0$, $\tilde{u}_0$, $v_0$ and $\phi_0$ in the framework of the covariant
reduction procedure, as is discussed in the next Section.
It will be shown that in the realization in question these superfields
are the only essential ones in terms of which all higher spin coset
superfields can be expressed after employing the inverse Higgs effect.

The expressions for other (super)forms are much more complicated and
it is not too enlightening to give them here.

Before ending this Section we mention that in order to obtain the Cartan
forms for the realizations with the stability subalgebras (\ref{StA1}) and
(\ref{StA0}), one should successively set equal to zero the coset
superfields $u_0$, $\tilde{u}_0$, $v_0$, $\phi_0$ and $\chi$, $\bar{\chi}$,
$u$, $v_1$, $v_2$.

\setcounter{equation}0

\section{$N=2$ Super Boussinesq equations from the covariant reduction}

In this section,
by applying the
inverse Higgs - covariant reduction procedure to the $sW_3^{\infty}$ coset
supermanifolds defined in the previous section,
we shall derive the evolution equations ($N=2$ super
Boussinesq equations) for a few essential coset superfields.
We will find the complete agreement with the Hamiltonian formulation
of ref. \cite{IK}. We
will also show that the spin 1 and spin 2 supercurrents, the
basic ingredients of $N=2$ super-$W_3$, naturally come out
in the nonlinear realization scheme as some coset parameters, in a close
analogy with the bosonic $W_3$ case \cite{u32a}. The $N=2$
super Miura maps \cite{{IK},{NY}} are then recognized as a part of the
inverse Higgs algebraic constraints covariantly relating these
parameter-superfields to the lower spin ones. To simplify the
presentaton,
here we choose the value of central charge $c$ to be 8. However,
all the subsequent equations can be easily
promoted to an arbitrary non-zero value of $c$ by a proper rescaling of
the involved superfields.

\subsection{Covariant reduction constraints}

As was explained in ref. \cite{{u32},{u32a}} and in Sec. 2, the basic idea
of the
covariant reduction is the imposition
of infinite number of covariant constraints on the initial
Cartan form $\Omega$, such that it is reduced to a one-form given
on an appropriate subalgebra (the covariant reduction subalgebra)
of the original (super)algebra. The necessary requirements the covariant
reduction subalgebra should obey [16-20] are: (i) it should contain the
stability subalgebra and (ii) it should
include the generators of (super)translations.

In order to encompass most general situation, we start with the Cartan
form $\Omega_{(2)}= g_{(2)}^{-1} d g_{(2)}$ corresponding to the realization
with the most narrow stability subalgebra (\ref{StA2}). We will perform the
covariant reduction of $\Omega_{(2)}$ successively, step by step,
first to the subalgebra $s\widetilde{{\cal H}}$ (\ref{CRA0}) and then to
two other covariant reduction subalgebras, $s\widetilde{{\cal H}}_{(1)}$
and $s\widetilde{{\cal H}}_{(2)}$, which are contained in (\ref{CRA0}) and
correspond to the realizations with the stability subalgebras (\ref{StA1})
and (\ref{StA2}), respectively. This chain of reductions can be expressed
as follows
\begin{eqnarray}
&\mbox{A}.& \Omega_{(2)} \Rightarrow \widetilde{\Omega}^{red} \in
s\widetilde {{\cal H}} \subset sW_3^{\infty} \label{CR} \\
&\mbox{B}.& \widetilde{\Omega}^{red} \Rightarrow
\widetilde{\Omega}_{(1)}^{red} \in
s\widetilde {{\cal H}}_{(1)} \subset s\widetilde {{\cal H}} \label{CR1} \\
&\mbox{C}.& \widetilde{\Omega}_{(1)}^{red} \Rightarrow
\widetilde{\Omega}_{(2)}^{red} \in
s\widetilde {{\cal H}}_{(2)} \subset s\widetilde {{\cal H}}_{(1)}\;.
\label{CR2}
\end{eqnarray}
The covariant reduction subalgebras $s\widetilde{{\cal H}}_{(1)}$ and
$s\widetilde{{\cal H}}_{(2)}$ are constituted by the
following sets of generators
\begin{equation} \label{CRA1}
s\widetilde {{\cal H}}_{(1)} = \left\{ s{\cal H}_{(1)},
W_{-2}, L_{-1}, G_{-1/2},\bar{G}_{-1/2} \right\}\;,
\end{equation}
\begin{equation} \label{CRA2}
s\widetilde{{\cal H}}_{(2)} =
\{\; s{\cal H}_{(2)}, W_{-2}, L_{-1}, G_{-1/2},\bar{G}_{-1/2} \}\;,
\end{equation}
where $s{\cal H}_{(1)}$ and $s{\cal H}_{(2)}$ have been defined in
eqs. (\ref{StA1}) and (\ref{StA2}). It is easy to check that
these generators indeed form closed sets. Recall that (\ref{CRA1}) is
the minimally possible (in the
framework of the ``contact'' $sW_3^{\infty}$ we are dealing with) closed
extension of $s{\cal H}_{(1)}$ which incorporates the (super)translations
generators
(see remark after eq. (\ref{StA1})).
It is also
worth mentioning that $s\widetilde{{\cal H}}_{(1)}$ and
$s\widetilde{{\cal H}}_{(2)}$ contain pieces of the
wedge superalgebra $sW_{\wedge}$ with some of the generators of
the quotient algebra
$sl(3|2)$, eq. (\ref{osp}). These sets of the $sl(3|2)$ generators
are closed modulo an ideal consisting of all the $s \geq 5/2$ higher spin
generators and so constitute quotients of
$s\widetilde{{\cal H}}_{(1)}$  and $s\widetilde{{\cal H}}_{(2)}$ by this
infinite-dimensional ideal
\begin{eqnarray}
sl_{(1)}(3|2) \sim  s\widetilde{{\cal H}}_{(1)}/
\{B^{(2)}_n,V_r^{(5/2)},\bar{V}_r^{(5/2)},\ldots  \} &=&
\{ J_0, L_0, \tilde{L}_0, W_0, L_{-1},  \tilde{L}_{-1},
W_{-2}, W_{-1}, G_{-1/2},  \nonumber \\
&& \bar{G}_{-1/2}, U_{-1/2}, \bar{U}_{-1/2},
U_{-3/2}, \bar{U}_{-3/2} \}\;, \label{sl1} \\
sl_{(2)}(3|2) \sim  s\widetilde{{\cal H}}_{(2)}/
\{B^{(2)}_n,V_r^{(5/2)},\bar{V}_r^{(5/2)},\ldots \} &=&
\{ L_{-1},  \tilde{L}_{-1},
W_{-2}, W_{-1}, G_{-1/2}, \bar{G}_{-1/2}, U_{-1/2},
\nonumber \\
&& \bar{U}_{-1/2}, U_{-3/2}, \bar{U}_{-3/2} \}\;.\label{sl2}
\end{eqnarray}

The equations (\ref{CR}) - (\ref{CR2}) are a concise notation for an
infinite sequence of constraints which follow from equating to zero
appropriate coefficients
in the decomposition of the relevant $\Omega$'s over the set of the
$sW_3^{\infty}$ generators. At the step A one equates to
zero all the parts of $\Omega_{(2)}$ which lie out of the subalgebra
$s\widetilde{{\cal H}}$; at the step B there appear additional constraints
stating that the components of $\Omega^{red}$ associated with the
generators which do not belong to $s\widetilde{{\cal H}}_{(1)} \subset
s\widetilde{{\cal H}}$ are zero; finally, at the step C, one puts equal to
zero also those components of $\Omega_{(1)}^{red}$ which are out of
$s\widetilde{{\cal H}}_{(2)} \subset s\widetilde{{\cal H}}_{(1)} \subset
s\widetilde{{\cal H}}$. Two types of constraints emerge: kinematical
(or algebraic) and
dynamical. The former constraints are just akin to
the inverse Higgs effect as it was originally formulated in \cite{IH},
they furnish covariant expressions for the higher spin coset superfields in
terms of
a finite number of the essential coset superfields and also imply some
irreducibility conditions for the latter. On the other hand, the
dynamical constraints lead to the dynamical
equations for the essential superfields. The covariance of the whole set
of constraints is guaranteed by the fact that all the component one-forms
equated to zero belong to the coset and so transform homogeneously
under $sW_3^{\infty}$, through each other. Note that the varieties of
constraints successively obtained by accomplishing the steps
(\ref{CR}) - (\ref{CR2}) are covariant in their own right.

\subsection{Expressing higher spin coset superfields}

One should keep in mind that
the vanishing
of any component coset one-form gives rise to three independent
equations for its
projections on the differentials $dt$, $d\theta$ and $d\bar{\theta}$
(the projection on $\Delta x$ is always expressed through the $d\theta$-
and $d\bar{\theta}$- projections by eq. (\ref{MC1})). The
equations for the $d\theta$ and $d\bar{\theta}$ projections basically
produce algebraic constraints while the equations
for the $dt$ projections yield the dynamics. Here we will exhaust the
algebraic consequences; the dynamical ones will be discussed in the next
Subsection.

A thorough inspection based on general arguments of ref. \cite{IH} and
on our previous experience of working with this kind of
nonlinear realizations shows that at the step A the only
independent coset
superfields which remain after solving the algebraic part of
constraints (\ref{CR}) are the spins 1 and 2 real superfields
$\phi(t,Z)$ and $\tilde{u}(t,Z)$ associated, respectively,
with the
generators $J_1$ and $\tilde{L}_2$. For instance, from the constraints
\begin{equation}
(j_{1})_{\theta} = (j_{1})_{\bar{\theta}} = (\tilde{l}_2)_{\theta} =
(\tilde{l}_2)_{\bar{\theta}} = 0
\end{equation}
one expresses the spin $3/2$ and spin $5/2$ coset superfields
$\mu$, $\bar{\mu}$  and $\nu_1$, $\bar{\nu}_1$
\begin{eqnarray}
\mu  =  \Db \phi & , & \bar\mu = -\D \phi \\
\nu_1 = \frac{1}{2} \Db {\tilde u }_2 & , &
{\bar\nu}_1 = -\frac{1}{2} \D {\tilde u }_2\;, \nonumber
\end{eqnarray}
{} from the constraints
\begin{equation}
(g_{3/2})_{\theta} = (g_{3/2})_{\bar{\theta}} = 0
\end{equation}
and their conjugate one expresses the spin 2 coset superfield $u_2$, etc.
It is easy to see that these two basic coset superfields are none other
than the basic $N=2$ super-$W_3$ supercurrents $J$ and $T$ defined in
Sec. 3 (up to unessential numerical rescalings). Indeed, $\phi$ and
$\tilde{u}_2$ are the only coset superfields which are shifted, respectively,
under the action of generators $J_1$ and $\tilde{L}_2$.
On the other hand, the $N=2$ super-$W_3$ transformations of $J$ and $T$
start with precisely the same
constant shifts. Actually, one might find the full transformation laws
of $\phi$ and $\tilde{u}_2$ and prove that these coincide with the
$N=2$ super-$W_3$ laws of $J$ and $T$. However, it is simpler to show
that for $\phi$ and $\tilde{u}_2$ in the present approach there arise the
same evolution equations and Miura maps as those for $J$ and $T$ in ref.
\cite{IK}.

To this end, let us continue the covariant reduction process and
switch on the step B constraints (\ref{CR1}).
At this level we find that the only independent superfields through which all
other coset parameters (including $\phi$ and $\tilde{u}_2$) can be
expressed are the complex spin $1/2$ superfields $\xi$ ($\bar{\xi}$) and
$\chi$ ($\bar{\chi}$) associated with the generators $G_{1/2}$
($\bar{G}_{1/2}$) and $U_{1/2}$ ($\bar{U}_{1/2}$). For instance, from
the constraints
\begin{equation}
(g_{1/2})_{\theta} = (g_{1/2})_{\bar{\theta}} = 0\;,\;\;\;\;
(f_{1/2})_{\theta} = (f_{1/2})_{\bar{\theta}} = 0
\end{equation}
the following expressions emerge for the first few coset superfields:
\begin{eqnarray}
u &=& \frac{1}{2}\left( {\cal D}\chi+{\bar {\cal D}}{\bar \chi}\right), \;\;\;
{\tilde u}=\frac{1}{4}\left({\cal D}\xi-{\bar {\cal D}}{\bar \xi}+
\chi{\bar \xi}-{\bar \chi} \xi \right)\;, \nonumber \\
v_1 &=& \frac{1}{6}\left( \xi{\bar \xi}+{\bar \chi}\xi+
\chi{\bar \xi}-{\cal D}\xi-{\bar {\cal D}}{\bar \xi}\right)\;, \label{HP} \\
\phi &=& \frac{1}{2}\left( {\bar {\cal D}}{\bar \chi}-
{\cal D}\chi-\xi{\bar \xi}-\chi{\bar \chi} \right)\;. \label{MM1}
\end{eqnarray}
Furthermore, the superfields $\chi$, $\bar{\chi}$, $\xi$, $\bar{\xi}$
turn out to be chiral (anti-chiral) because, also as a
consequence of constraints (5.11), we obtain the
following irreducibility conditions
\footnote{
Analogous chirality conditions appeared in ref. \cite{u31a} in the process
of deducing $N=2$ super Liouville equation within the covariant reduction
procedure applied to $N=2$ superconformal algebra.}
for these superfields:
\begin{equation} \label{Ch}
{\cal D}{\bar \chi}={\bar {\cal D}}\chi={\cal D}{\bar \xi}=
{\bar {\cal D}}\xi=0\;.
\end{equation}

The expressions for
other higher spin coset superfields are much more complicated, though
all of them can be straightforwardly computed from the conditions of
vanishing of the appropriate higher order component one-forms contained
in $\Omega$. All these superfields are eventually expressed in terms of
$\xi$ and
$\chi$. In particular, as one of the consequences of the constraint
\begin{equation}
\tilde{l}_1 = 0\;,
\end{equation}
one obtains the expression for $\tilde{u}_2$
\begin{eqnarray}
&&12\; \tilde{u}_2 = \partial ({\cal D}\xi-{\bar {\cal D}}{\bar \xi})+
2\partial \xi {\bar \chi}+
\partial \xi{\bar \xi}+\partial {\bar \xi}\xi-2 \partial {\bar \xi}\chi+
\partial \chi {\bar\xi}- \partial{\bar \chi}\xi-5 \xi{\bar \xi}\chi{\bar \chi}
\nonumber\\
&&\quad +{\bar {\cal D}}{\bar \xi}\Bigl[{\bar {\cal D}}{\bar \xi}-
{\cal D}\chi-{\bar {\cal D}}{\bar \chi}+
\chi{\bar \chi}-\xi{\bar \xi}+2{\bar \xi}\chi-4\xi{\bar \chi}\Bigr]
\nonumber \\
&&\quad +{\cal D}\xi\Bigl[{\cal D}\chi+{\bar {\cal D}}{\bar \chi}+
{\cal D}\xi-3{\bar {\cal D}}{\bar \xi}+
2\xi{\bar \chi}-4{\bar \xi}\chi+\chi{\bar \chi}-\xi{\bar \xi}\Bigr]
\nonumber\\
&&\quad +{\cal D}\chi\Bigl[\xi{\bar \chi}-2{\bar \xi}\chi -\xi{\bar \xi}\Bigr]+
{\bar {\cal D}}{\bar \chi}\Bigl[2\xi{\bar \chi}-{\bar \xi}\chi+
\xi{\bar \xi}\Bigr].
\label{MM2}
\end{eqnarray}
Comparing (\ref{MM1}) and (\ref{MM2}) with
the super Miura maps
relating the $N=2$  $W_3$ supercurrents $J$ and $T$ to two spin
$1/2$ chiral supercurrents \cite{IK} we observe the
complete coincidence between them. This confirms the
identification of $\phi$ and $\tilde{u}_2$ as the $N=2$ $W_3$ supercurrents
\begin{equation} \label{Id}
T \;\equiv\; 12 \tilde{u}_2\;,\;\;\;\;\; J \;\equiv\; 2 \phi\;,
\end{equation}
and suggests that $\chi$ and $\xi$ can be identified with the
above spin $1/2$ supercurrents.

For our further purposes we will need, besides the expressions already
presented, explicit expressions only for the coset superfields
$u_2,v_2,\phi_2$ entering the functions
$(l_0)_t,({\tilde l}_0)_t, (w_0)_t$ and $(j_0)_t$ in eqs.(\ref{f12}).
These expressions are given in eq. (C.8-C.10 ). They are
obtained  from the constraints
\begin{equation}
(g_{3/2})_{\theta} = (g_{3/2})_{\bar{\theta}} =0\;, \;\;\;\;\;
(f_{3/2})_{\theta} = (f_{3/2})_{\bar{\theta}} =0\;.
\end{equation}
We do not quote explicit expressions for the Cartan forms $g_{3/2}$ and
$f_{3/2}$ in view of their complexity. Note that at the step B the
spin 0 coset superfields $u_0, \tilde{u}_0, \phi_0$ and $v_0$ are pure
gauge and so can be completely gauged away: the constraints (\ref{CR}),
(\ref{CR1}) are covariant under
arbitrary right gauge shifts of $g_{(2)}$ (4.10) by the elements
of the stability subgroup $sH_{(1)}$ associated with the algebra (\ref{StA1}).

At the step C (eq. (\ref{CR2})) this gauge invariance gets broken down to
the invariance with respect to right $sH_{(2)}$ multiplications, so
the scalar
coset superfields just
mentioned cease to be pure gauge: on the contrary, they become the
essential superfields in this case. The set of the step C constraints
incorporates all the constraints imposed before and contains four
additional ones
\begin{equation} \label{Exte}
l_0 = {\tilde l}_0 = w_0 = j_0 =0 \;,
\end{equation}
where the one-forms $j_0, l_0, \tilde{l}_0$ and $w_0$ were defined in eq.
(\ref{f12}). The $d\theta$ and $d\bar{\theta}$ projections of these new
constraints express the spin 1/2 coset superfields $\chi$ and $\xi$ in terms
of the spin 0 ones
\begin{equation} \label{120}
\chi = -\Db {\overline\Phi} \; , \;
\bar\chi = \D \Phi \; , \;
\xi = - \Db {\overline{\cal V}} \; , \;
\bar\xi =  \D {\cal V} \; ,
\end{equation}
and simultaneously imply the chirality conditions for the latter
\begin{equation} \label{Ch1}
\Db\Phi = \D \overline{\Phi} =0 \;, \;  \Db{\cal V} =
\D \overline{\cal V} =0\;.
\end{equation}
Here
\begin{equation}
\Phi = -\left( \phi_0+\frac{1}{2}u_0\right) \; ,\;
\overline{\Phi} = -\left( \phi_0-\frac{1}{2}u_0\right)\; ,\;
{\cal V} = v_0+{\tilde u}_0\;,\;
\overline{\cal V} = -v_0+{\tilde u}_0\;.
\end{equation}
Thus, in this case the only essential coset parameter-superfields are
two complex chiral spin 0 superfields $\Phi$ and ${\cal V}$. After
substitution of (\ref{120}) into (\ref{MM1}) and (\ref{MM2}) one
recognizes the latter as the Miura maps of the $N=2$ $W_3$ supercurrents
onto the scalar chiral $N=2$ superfields \cite{{NY},{IK}}.

The main conclusion following from the above consideration is that the whole
$sW_3^{\infty}$ symmetry can be realized (in a nonlinear way) on the
finite-dimensional manifolds:
\begin{equation} \label{R1}
\{t, Z \equiv (x, \theta, \bar{\theta}),\; \chi (t,Z),\; \bar{\chi} (t,Z),\;
\xi (t,Z),\;\bar{\xi} (t,Z) \}\;,
\end{equation}
and
\begin{equation} \label{R2}
\{ t, Z , \Phi (t,Z),\; \bar{\Phi} (t,Z),\;
{\cal V} (t,Z),\;\bar{{\cal V}} (t,Z) \}\;,
\end{equation}
where the coordinates-superfields $\chi,\;\xi$ and $\Phi, \;{\cal V}$ satisfy
the chirality conditions (\ref{Ch}) and (\ref{Ch1}). Let us point out once
again that the $N=2$ superspace
coordinates $(t, Z)$ do not form an invariant subspace in (\ref{R1}) and
(\ref{R2}): the $sW_3^{\infty}$ transformations mix them with the
coordinates-superfields and derivatives of the latter
(of all orders). We also note that the closure of $sW_3^{\infty}$ on these
sets is achieved
only by making use of the evolution ($N=2$ Boussinesq) equations for
the involved superfields (see next Subsection) because the algebraic and
dynamical parts of the above covariant reduction constraints are mixed under
the $sW_3^{\infty}$ transformations (like in the $W_3$ case
\cite{{u32},{u32a}}).

This is an appropriate place to summarize the group-theoretical and geometric
meaning of these realizations and to compare them with the analogous
realizations of $W_3^{\infty}$ \cite{u32a}.

The sets (\ref{R1}), (\ref{R2})
define the covariant embeddings of $N=2$ superspace $(t, Z)$ into the cosets
$sW_3^{\infty}/ sH_{(1)}$ and $sW_3^{\infty}/ sH_{(2)}$, where the
supergroups in the denominator are related to the
superalgebras (\ref{StA1}) and (\ref{StA2}), respectively. These embeddings
are
fully specified by the superfields $\chi, \xi$ and $\Phi, {\cal V}$ which are
subjected to the evolution equations to be given below. Note that in both
cases the superfields can be regarded as the essential parameters
of the cosets of $sW_3^{\infty}$ over the covariant reduction
subgroups, i.e. the subgroups with the algebras
$s\widetilde{{\cal H}}_{(1)}$ and
$s\widetilde{{\cal H}}_{(2)}$. The $N=2$, $2D$ superspace itself
can be identified with the coset of the supergroups $SL_{(1)}(2|3)$
and $SL_{(2)}(2|3)$ (eqs.
(\ref{sl1}), (\ref{sl2})) over their subgroups generated by
$$
\{ W_0, L_0, \widetilde{L}_0, J_0, U_{-3/2}, \bar{U}_{-3/2},
G_{-1/2}+U_{-1/2}, \bar{G}_{-1/2}-\bar{U}_{-1/2},
L_{-1}-\widetilde{L}_{-1}, W_{-1}+\frac{3}{2}L_{-1} \}
$$
and
$$
\{ U_{-3/2}, \bar{U}_{-3/2},
G_{-1/2}+U_{-1/2}, \bar{G}_{-1/2}-\bar{U}_{-1/2},
L_{-1}-\widetilde{L}_{-1}, W_{-1}+\frac{3}{2}L_{-1} \}\;,
$$
respectively (recall that these sets are closed modulo higher spin
generators). Bosonic analogs of the supermanifolds (\ref{R1}) and
(\ref{R2}) are the manifolds
$$
(t,x, u_1(t,x), v_1(t,x))
$$
and
$$
(t,x, u_0(t,x), v_0(t,x))
$$
which are closed under the action of $W_3^{\infty}$
(see Sec.2 and ref.\cite{u32a}). The fields
$u_1, v_1$ and $u_0, v_0$ are the essential parameters of the cosets of
$W_3^{\infty}$ over the subgroups associated with the algebras
$\widetilde{{\cal H}}_{(1)}$ and $\widetilde{{\cal H}}_{(2)}$, while the
space-time coordinates  $(t,x)$ in both cases parametrize
the cosets of the latter
subgroups over the relevant stability subgroups. Equivalently, the
coordinates $(t, x)$
can be viewed as parametrizing the cosets of appropriate subgroups of the
quotient group $SL(3,R)$ with the algebra (\ref{sl}).

In the bosonic case there is one more invariant manifold, namely
$(t,x, u_2(t,x), v_3(t,x))$, with $u_2$ and $v_3$ being the spin 2 and
spin 3 $W_3$ currents \cite{u32a}. The basic reason why this manifold
is closed under $W_3^{\infty}$ is that it still admits a coset interpretation:
the currents are essential parameters
of the coset $W_3^{\infty}/ \widetilde{H}$, with  generated
by $\widetilde{{\cal H}} = \{ sl(3,R) + \mbox{Higher spin generators}
\}$, whereas $t, x$ are parameters of a
two-dimensional coset of $\widetilde{H}$, with $W_{-2}$ and $L_{-1}$
as the coset generators.

In the supersymmetric case, despite the fact that the covariant reduction
(5.1) leaves the supercurrents $\phi, \tilde{u}_2$ as the only
essential parameters of the coset $sW_3^{\infty}/ s\widetilde{H}$, the set
\begin{equation} \label{R3}
\{ t, Z, \phi(t,Z), \tilde{u}_2 (t,Z) \}
\end{equation}
is not closed under the left action of $sW_3^{\infty}$. The reason
has been already mentioned in Subsec. 4.2 and it consists in that the $N=2$
superspace $(t, Z)$ cannot be regarded as a coset manifold of
the supergroup $s\widetilde{H}$: such a manifold of minimal dimension
contains, besides the
$N=2$ superspace coordinates, also the coset superfield $\tilde{u}$ and a
linear combination of $\xi$ and $\chi$, which
follows from comparing eqs. (\ref{CRA0}) and (\ref{StA0}). Thus,
only
at cost of adding at least
this minimal number of extra superfields,
the set (\ref{R3}) can be promoted to an
invariant space of $sW_3^{\infty}$. Note, however, that the
constraints (\ref{CR1})
are covariant with respect to the $SL(3|2)$ gauge transformations realized
as right $SL(3|2)$ shifts of the coset element (\ref{CE}) with arbitrary
parameter-superfunctions. Using this freedom, one may choose the gauge so
as to kill all the additional superfields mentioned above.
In other words, the set (\ref{R3}) is
invariant under modified $sW_3^{\infty}$ transformations which are closed
modulo a compensating gauge $SL(3|2)$ transformation. Here we will not dwell
more on this point.

\subsection{Dynamics}

As was already noticed, the equations restricting the $t$ dependence of the
essential coset superfields come from the $dt$ projections of the covariant
reduction constraints. The dynamical constraints arising at the steps
A - C (eqs.(\ref{CR}) - (\ref{CR2})) are as follows
\begin{eqnarray}
& A.&  (j_1)_t = (\tilde{l}_2)_t = 0\;, \label{BR} \\
& B.&  (g_{1/2})_t = (f_{1/2})_t = 0 \;, \label{BR1}\\
& C.&  (l_0)_t = (\tilde{l}_0)_t = (j_0)_t = (w_0)_t = 0\;. \label{BR2}
\end{eqnarray}
After substituting the inverse Higgs expressions for the involved
higher spin coset superfields, these constraints yield the evolution
equations for the relevant pairs of the essential
superfields: $\phi$ and $\tilde{u}_2$, $\chi$ and $\xi$, $\Phi$ and
${\cal V}$, respectively. Only taking account of
these evolution equations the sets (\ref{R1}) and (\ref{R2}) are
actually closed under $sW_3^{\infty}$ symmetry.

Before presenting the explicit form of the equations, let us recall
that up to now, for simplicity, we  assumed that the $t$
translation generator is $W_{-2}$. However, it is desirable to
consider
more general situation, identifying the generator of $t$ translations with
$W_{-2}^{(\alpha)}$ defined in eq. (\ref{Wnew}) and placing the latter
in the coset element (\ref{CE}) instead of $W_{-2}$. Thereupon, most of
formulas actually needed for our purposes, undergo only slight modifications:
e.g.,  in the
expressions (\ref{f12}) the dependence on the parameter
$\alpha$ appears
only in the functions
$(l_0)_t,({\tilde l}_0)_t,(w_0)_t$ and $(j_0)_t$
(given by eqs. (C.1-C.4 ) of Appendix C).
Using these modified expressions, it is straightforward to find from
(\ref{BR}), (\ref{BR1}) and (\ref{BR2}) the general form of the sought
pairs of the
evolution equations
\begin{eqnarray}
{\dot T} \;\equiv \; 12\dot{\tilde{u}}_2 &=& 2 J'''-\left[\Db ,\D\right]  T'-
 10\partial\left( \Db J \D J\right)
 +4J'\left[\Db , \D\right]J +2J\left[\Db , \D\right]J'
-4J^2J' \nonumber \\
 & & -\left( 5-\alpha \right)\Db J\D  T
     -\left( 5-\alpha \right)\D J\Db  T
 -\left(8+2\alpha\right)J' T
  -\left(3+\alpha\right)J T' \label{SB0} \\
{\dot J} \;\equiv \; 2\dot{\phi} &=& -2 T'-\alpha\left(\left[\Db ,\D\right]J'
 +2JJ'\right)  \quad, \label{SB00}
\end{eqnarray}
\begin{eqnarray} \label{SB1}
&&{\dot \chi}=2{\partial}^2\xi-\alpha{\partial}^2\chi+
(5-\alpha)\Db\D(\chi\xi{\bar \xi})+
\partial \Bigl[ 2\Db{\bar \xi}\chi-2\alpha\Db{\bar \chi}
\chi-(3+\alpha)\Db{\bar \xi}\xi+4\Db{\bar \chi}\xi\Bigr]
\nonumber \\
&& \qquad +\partial\chi\Bigl[2{\cal D}\xi+2\alpha{\cal D}\chi\Bigr]+
\partial\xi\Bigl[2{\cal D}\chi+4{\cal D}\xi\Bigr] \quad ,
\end{eqnarray}
\begin{eqnarray} \label{SB2}
&&{\dot \xi}=-{\partial}^2\xi-2{\partial}^2\chi+
(5-\alpha )\Db\D ( \chi{\bar \chi}\xi)+
\partial \Bigl[2\Db{\bar \chi}\chi-2\Db\bar\xi\xi-
(3+\alpha )\Db{\bar \chi}\xi-4\Db{\bar \xi}\chi\Bigr]
\nonumber\\
&& \qquad +\partial\chi\Bigl[4{\cal D}\chi+(3+\alpha){\cal D}\xi\Bigr]+
\partial\xi\Bigl[(3+\alpha ){\cal D}\chi-2\Db{\bar \xi}\Bigr]\quad,
\end{eqnarray}
\begin{eqnarray}
{\dot \Phi} &=& 2\partial^2{\cal V} + \alpha \partial^2 \Phi +
 2 \D{\Phi}\partial\Db\overline{\cal V}+
2\alpha\D\Phi\partial\Db\overline{\Phi}+
 4 \D{\cal V}\partial\Db\overline{\Phi} +
  (3+\alpha)\D{\cal V}\partial\Db\overline{{\cal V}} +
  2\partial\Phi\partial\overline{{\cal V}} \nonumber \\
&&+ 4\partial{\cal V}\partial\overline{\Phi}
-2\partial{\cal V}\partial{\Phi}+2\partial{\cal V}\partial{\cal V}+
 \alpha(2\partial\Phi\partial\overline{\Phi}+\partial\Phi\partial\Phi )+
(3+\alpha )\partial{\cal V}\partial\overline{{\cal V}} \nonumber \\
&& +(5-\alpha)\left( \partial{\cal V}\D\Phi\Db\overline{\cal V} -
  \D{\cal V}\partial\Phi\Db\overline{{\cal V}} \right) \quad, \label{SB3}
\end{eqnarray}
\begin{eqnarray}
 \dot{\cal V} &=& \partial^2 {\cal V}-2\partial^2 \Phi +
 2\D\Phi\partial\Db\overline{\Phi}+4\D\Phi\partial\Db\overline{{\cal V}}+
(3+\alpha)(\D{\cal V}\partial\Db\overline{\Phi}+
 \partial{\cal V}\partial\overline{\Phi}+\partial{\cal V}\partial\Phi )
 \nonumber \\
&&- 2\D{\cal V}\partial\Db\overline{{\cal V}}+
2\partial\Phi\partial\overline{\Phi}+4\partial\Phi\partial\overline{{\cal V}}-
2\partial\Phi\partial\Phi -2\partial{\cal V}\partial\overline{{\cal V}}+
\partial{\cal V}\partial{\cal V} \nonumber \\
&&+(5-\alpha)(\D{\cal V}\partial\Phi\Db\overline{\Phi}-
\partial{\cal V}\D\Phi\Db\overline{\Phi}) \quad. \label{SB4}
\end{eqnarray}
The equations for ${\bar \xi}$ and ${\bar \chi}$ can be obtained from
eqs. (\ref{SB1}) and (\ref{SB2}) by applying the same rules as
for the
generators $G$ and $U$ in the Appendix B.

The system (\ref{SB0}), (\ref{SB00}) coincides with the
$N=2$ super Boussinesq equation
derived in \cite{IK} as a hamiltonian flow on $N=2$ super-$W_3$
(the detailed comparison with the hamiltonian approach will be given in
the next section)\footnote{The equations given in \cite{IK} contain
some misprints in the numerical coefficients.}.
Thus we conclude that
this equation can be alternatively derived in a pure geometric way as
one of the conditions of embedding the $N=2$ superspace $(t, Z)$ as a
geodesic supersurface into the infinite-dimensional coset supermanifold
$sW_3^{\infty}/s\widetilde{H}$. As the superfields $\chi, \xi, \Phi$ and
${\cal V}$ are related to $\phi, \tilde{u}_2$ via super Miura maps
(\ref{MM1}), (\ref{MM2}), (\ref{120}), the evolution equations for them
can naturally be called modified $N=2$ super Boussinesq equations. Their
geometric interpretation
within the coset space approach is quite similar to that of $N=2$ super
Boussinesq equation. It is straightforward to check that the
whole set of equations (\ref{SB0}) - (\ref{SB4}) is compatible
with the super Miura maps:  e.g., taking the $t$ derivative of both sides of
(\ref{120}) and using eqs. (\ref{SB3}), (\ref{SB4}), one gets for
$\chi, \xi$ just eqs. (\ref{SB1}), (\ref{SB2}), etc.

It is remarkable that all these three kinds of $N=2$ super Boussinesq
equation together with the relevant super Miura maps naturally come out
within the single geometric procedure: covariant reduction of the coset
supermanifolds of the $sW_3^{\infty}$ symmetry.

\setcounter{equation}0

\section{Comparison with the hamiltonian approach. Integrability
properties}

It is well-known that the $W_3$ algebra (2.1) provides the second
hamiltonian structure for the Boussinesq system (\ref{26a})
\cite{SCW}: the latter can be written in the Hamilton form
\begin{equation}
\dot{W} = \left\{ W, H \right\}_{PB}\;, \;\;\;
\dot{T} = \left\{ T, H \right\}_{PB}\;,
\end{equation}
where
$$
H \propto \int dx W (t,x)
$$
and $T(t,x), W(t,x)$ are the spin 2 and spin 3 currents which, at equal time,
satisfy the standard Poisson brackets (or, equivalently, OPE's) of the
$W_3$ algebra (2.1).

In a similar way, it
has been demonstrated in ref.\cite{IK} that for the $N=2$ super-$W_{3}$ case
the most general supersymmetric form of the hamiltonian is as follows:
\begin{equation} \label{HS}
H=- \int dZ \bigl(T+\frac{\alpha}{2} J^2 \bigr)\;,
\end{equation}
where $dZ \equiv dx d\theta d\bar\theta$. Assuming that the supercurrents
$J$ and $T$ obey the SOPE's of $N=2$ super-$W_3$ algebra as they are
given in Appendix A, the evolution equations associated with this
Hamiltonian,
\begin{equation}
\dot{J} = \left\{ J, H \right\}_{PB}\;, \;\;\; \dot{T} =
\left\{ T, H \right\}_{PB}\;,
\end{equation}
coincide with eqs.(\ref{SB0}), (\ref{SB00}) obtained in the
geometric framework of the coset space approach.

In order to establish a link between the two approaches, we expand the
integrand
in (\ref{HS}) in Laurent series in $x$ and integrate over $dx$ and $d\theta,
d\bar{\theta}$. As a result we obtain
\begin{equation} \label{HW}
H \propto W_{-2}+\alpha\bigl(TJ- G{\bar G}\bigr)_{-2}
= W_{-2}^{(\alpha)}\;,
\end{equation}
i.e. just the most general $t$ translations generator (\ref{Wnew}). Thus,
the freedom in the choice of this generator within the coset space approach
reflects the freedom in
the choice of the hamiltonian in the framework of the approach based on the
notion of the second hamiltonian structure.

It is straightforward to check that the evolution equation for the
spin $1/2$ chiral supercurrents $\chi$ and $\xi$ (eqs.(\ref{SB1}),
(\ref{SB2})) can be
written in the Hamilton form with the same hamiltonian (\ref{HS})
\begin{equation}
{\dot \chi}=\left\{\chi, H \right\}_{PB}\;, \;\;\;\;
{\dot \xi}=\left\{ \xi, H\right\}_{PB}\;,
\end{equation}
if one assumes that the two-point functions of these supercurrents are given
by the standard expressions characteristic of the spin-1/2 chiral
superfields, namely:
\begin{equation} \label{2psp}
<\chi(Z_1){\bar \chi}(Z_2)>={1 \over Z_{12}}+
{\theta_{12} \bar\theta_{12} \over 2 Z^2_{12}}\;,\;\;
<\xi(Z_1){\bar \xi}(Z_2)>={1 \over Z_{12}}+
{\theta_{12} \bar\theta_{12} \over 2 Z^2_{12}}
\end{equation}
(for notation see Appendix A).
Analogously, the equations (\ref{SB3}), (\ref{SB4})
can be recovered in the hamiltonian formalism assuming
that $\Phi$, ${\cal V}$ are free
chiral $N=2$ superfields \cite{IK}:
\begin{equation} \label{2pss}
<\Phi(Z_1){\bar \Phi}(Z_2)>= \mbox{ln} (Z_{12}) -
{\theta_{12} \bar\theta_{12} \over 2 Z_{12}}\;, \;\;
<{\cal V}(Z_1){\bar {\cal V}}(Z_2)>= \mbox{ln} (Z_{12}) -
{\theta_{12} \bar\theta_{12} \over 2 Z_{12}}\;.
\end{equation}
We also note that the SOPE's (\ref{2psp}), (\ref{2pss}) produce
for the Miura expressions (\ref{MM1}), (\ref{MM2}) of the supercurrents
precisely the SOPE's
((A.1)-(A.3)) of $N=2$ super-$W_3$ algebra.

Finally, let us briefly discuss the integrability issues.

As has been noticed in Sec. 2, the covariant reduction procedure
automatically yields
a zero curvature representaton for the dynamical equations obtained, which
is a consequence of the covariant reduction constraints and the Maurer-Cartan
identity for the original Cartan form. For the reduced Cartan forms arising
at different stages of the covariant reduction (eqs.
(\ref{CR}) - (\ref{CR2})) the zero curvature condition reads
\begin{equation}  \label{ZC}
d^{ext} \Omega^{red} = \Omega^{red} \wedge
\Omega^{red} \;,
\end{equation}
where $\Omega^{red}$ stands for $\widetilde{\Omega}^{red},\;
\widetilde{\Omega}^{red}_{(1)}, \; \widetilde{\Omega}^{red}_{(2)}$.
Since the higher spin generators form ideals in the relevant covariant
reduction subalgebras, those
parts of the reduced Cartan form which are valued in the quotient algebra
$sl(3|2)$ satisfy (\ref{ZC}) in their own right. So, without loss of
generality, we may keep in $\Omega^{red}$ only these parts
\begin{equation} \label{Val}
\widetilde{\Omega}^{red} \in sl(3|2)\;, \;\;
\widetilde{\Omega}^{red}_{(1)} \in sl_{(1)}(3|2)\;,\;\;
\widetilde{\Omega}^{red}_{(2)} \in sl_{(2)}(3|2)\;,
\end{equation}
where the superalgebras in the r.h.s. were defined by
eqs. (\ref{osp}), (\ref{sl1}), (\ref{sl2}).
Here we will not give these Cartan forms explicitly, though they can be
readily evaluated; let us only recall that, using the
appropriate gauge freedom indicated in Subsec. 5.2, all these forms can be
written entirely in terms of the relevant essential superfields:
$\widetilde{\Omega}^{red}$ via $J$ and $T$, $\widetilde{\Omega}^{red}_{(1)}$
via $\chi,\; \bar{\chi}, \; \xi, \;\bar{\xi}$ and
$\widetilde{\Omega}^{red}_{(2)}$
via $\Phi, \;\bar{\Phi},\; {\cal V}, \;\bar{\cal V}$.

Recall also that not all of the projections of $\Omega^{red}$ on the
differentials $dt, \Delta x, d\theta, d\bar{\theta}$ are independent, the
$\Delta x$ projection is expressed through spinor ones by eq. (\ref{MC1}).
For completeness, we quote here all the independent projections of the zero
curvature condition (\ref{ZC})
\begin{eqnarray} \label{ZC1}
&& {\cal D}\Omega^{red}_t + \dot{\Omega}^{red}_{\theta} +
[ \Omega^{red}_t, \Omega^{red}_{\theta} ] = 0\;, \nonumber \\
&& \bar{{\cal D}}\Omega^{red}_t + \dot{\Omega}^{red}_{\bar{\theta}} +
[ \Omega^{red}_t, \Omega^{red}_{\bar{\theta}} ] = 0\;, \nonumber \\
&& {\cal D} \Omega^{red}_{\theta} - \{\Omega^{red}_{\theta},
\Omega^{red}_{\theta}\} = 0 \nonumber \\
&& \bar{{\cal D}} \Omega^{red}_{\bar{\theta}} - \{\Omega^{red}_{\bar{\theta}},
\Omega^{red}_{\bar{\theta}}\} = 0 \;.
\end{eqnarray}
After substitution of the appropriate expressions for the involved
$sl(3|2)$ valued projections of $\Omega^{red}$, these relations
yield the evolution equations (\ref{SB0}) - (\ref{SB4}).

We point out that the above zero curvature representation exists for any
choice of the parameter $\alpha$ in eqs. (\ref{SB0}) - (\ref{SB4}).
However, recently we have found \cite{LP} that the $N=2$ super Boussinesq
equation (\ref{SB0}), (\ref{SB00}) admits higher order conserved quantities
only for three selected values of $\alpha$, $\alpha = (-4,-1, 5) $.
This means
that it is integrable only in these special cases (cf. $N=2$ super KdV
equation \cite{SKdV}) despite the fact that it possesses a zero curvature
representation for any value of $\alpha$ (the same, of course, is true
for the modified $N=2$ super Boussinesq equations). Hence, the existence of
such a representation is a weaker requirement than integrability and
there arises the question how to understand the aforementioned
restrictions on $\alpha$ within the coset space approach.

The only conceivable answer seems to be as follows. In the hamiltonian
formalism the higher conserved quantities of the $N=2$ super Boussinesq
equation are the hamiltonians for the higher equations from the $N=2$
Boussinesq hierarchy. Their characteristic property is that they commute
with each other and with the basic hamiltonian (\ref{HS}).
After performing the
integration over the $N=2$ superspace coordinates $Z$ they should be
recognized as appropriate modes of some composite objects constructed
{}from the supercurrents $J$ and $T$, like $H$ (\ref{HS}) has been
recognized as the generator $W_{-2}^{(\alpha)}$. Hence, in the
$sW_3^{\infty}$ language,
searching for the higher order conserved quantities amounts to
singling out the sequences of the higher spin $sW_3^{\infty}$
generators which commute with the $t$ translation generator
$W_{-2}^{(\alpha)}$ and among themselves. Then the result of \cite{LP}
means that these infinite
sequences exist only for the three values of $\alpha$ indicated above.
In order to ensure the integrability in the above sense, i.e.
as the existence of infinitely many conserved quantities in involution,
one is led to
pass to another, more general realization of $sW_3^{\infty}$, with all
mutually commuting $t$ translations generators placed in the coset. This
will entail introducing infinitely many ``time''coordinates  and
allowing all coset superfields to depend on these coordinates. One may
expect that the covariant reduction will still express all these
superfields in terms of the two essential ones $J$ and $T$
and simultaneously yield evolution equations for the latter
with respect to each ``time'' coordinate. In this way the whole $N=2$ super
Boussinesq hierarchy could come out within the coset space approach.
In other words, we hope that the value of $\alpha$ can be properly
fixed in this approach if we will pose the problem of deducing
the whole $N=2$ super Boussinesq hierarchy rather than its first
nontrivial representative (\ref{SB0}), (\ref{SB00}). We will examine
this intriguing possibility in more detail elsewhere.

\section{Conclusion}

The key ingredient in our study of $W_3$ and $N=2$ super-$W_{3}$ symmetries
in the framework of the method of coset space realizations
is the construction of linear
infinite-dimensional algebras $W_3^{\infty}$
and $sW_{3}^{\infty}$ from the standard
nonlinear $W_3$ and $N=2$ super-$W_{3}$ algebras by treating as independent
generators the Laurent modes of all the composite currents present in
the enveloping algebra of the basic currents of these algebras
\cite{{u32},{u32a}}.
The application of the standard techniques of nonlinear realizations
augmented with the ideas of the inverse Higgs effect and the covariant
reduction lead to the interpretation of the plane $t, x$ and its
superextension, $N=2$ superspace $t, x, \theta, \bar{\theta}$, as geodesic
submanifolds embedded into infinite-dimensional coset manifolds of
$W_3^{\infty}$ and $sW_3^{\infty}$ symmetries. These embeddings are
completely specified by finite numbers of the essential coset
parameter-(super)fields which are recognized either as the basic
(super)currents generating the original nonlinear algebras or as the lower
spin (super)fields related to the former ones via (super)Miura maps.
The Miura maps together with the evolution equations for the essential
(super)fields (Boussinesq and $N=2$ super Boussinesq equations as well as
their modified versions) naturally come out as the most essential part of
the embedding conditions. The $W_3$ and $N=2$ super-$W_3$ symmetries turn
out to be the particular realizations of $W_3^{\infty}$ and $sW_3^{\infty}$
preserving the embeddings just mentioned. The characteristic feature of
these realizations is that they necessarily mix the coordinates $t, x$ or
$t, x, \theta, \bar{\theta}$ with the (super)functions specifying the
embedding.

By exploiting the ideas of the coset space approach we have been thus able to
provide a common geometric basis to the following $2D$ integrable systems:
the Liouville and super Liouville equations \cite{{u30},{u31},{u31a}}, the
$sl_3$ Toda equations \cite{u32}, the Boussinesq and modified
Boussinesq equations \cite{u32a}, the $N=2$ super Boussinesq equations.
We are sure that a variety of other $2D$ integrable systems, at least those
respecting conformal invariance, can be described on similar grounds, by
applying the covariant reduction to other nonlinear $W$ type
algebras and superalgebras. Besides new integrable systems and
various free-field type representations for them (via the relevant Miura
maps), we expect to obtain in this way new intrinsic relations between
different systems related to the same $W$ algebra, e.g., between Boussinesq
equations and the $sl_3$ Toda equations. An interesting problem is to treat
the full quantum $W$ algebras in the same language of passing to
linear $W^{\infty}$ type algebras and to work out convenient
geometric methods for deducing associated quantum evolution equations.
It would be also of interest to find out
possible implications of our geometric approach in $W$ strings, $W$ gravity
and related theories.

\setcounter{equation}{0}
\vspace{1.5cm}

{\Large\bf Appendix A}
\def\theequation{A.\arabic{equation}}

\begin{center}
{\large\bf $N=2$ super-$W_3$ algebra: SOPE's and OPE's}
\end{center} \vspace{0.3cm}

In this Appendix we present the formulations of the classical
$N=2$ super-$W_3$ algebra in terms of SOPE's of the supercurrents
$J(Z)$, $T(Z)$ and OPE's of the component currents.

The full set of the relevant SOPE's reads

\begin{equation}\label{1}
\jl\jr=\frac{c}{4 \Z^2}+ \frac{\Tb\Db J}{\Z}-\frac{\T\D J}{\Z}
+\frac{\T\Tb J}{\Z^2}+\frac{\T\Tb\partial J}{\Z} \; ,
\end{equation}
\begin{equation}\label{2}
\jl\tr= \frac{\Tb\Db T}{\Z}-\frac{\T\D  T}{\Z}
+2\frac{\T\Tb T}{\Z^2}+\frac{\T\Tb\partial T}{\Z} \; ,
\end{equation}

$$
\tl\tr =  -\frac{3c}{2\Z^4}-12\frac{\T\Tb J}{\Z^4}+
12\frac{\T\D J}{\Z^3}-12\frac{\Tb\Db J}{\Z^3}-
12\frac{\T\Tb\partial J}{\Z^3}
$$
$$
+ 2\frac{5 T -2\left[\Db,\D\right] J + B^{(2)} }{\Z^2}
+ \frac{\T\D\left(  8\partial J + 5T + B^{(2)}\right)}{\Z^2}
$$
$$
-\frac{\Tb\Db\left(  8\partial J - 5T - B^{(2)} \right)}{\Z^2}
 +  \frac{\T\Tb\left({3\over 2}\left[\Db,\D\right] T-
    6\partial^2 J+U^{(3)}\right)}{\Z^2}
$$
$$
 + \frac{\T\left( 3\partial\D T
 +3\partial^2\D J+{\overline\Psi}^{(7/ 2)}
     \right)}{\Z}
    -\frac{\Tb\left(-3\partial\Db T+3\partial^2\Db J
    +\Psi^{(7/2)}  \right)}{\Z}
$$
$$
+ \frac{\T\Tb}{\Z}\left( -2\partial^3 J+\partial\left[ \Db,\D\right] T
 + \frac{1}{2}\partial U^{(3)} +\frac{1}{2}\Db{\overline\Psi}^{(7/2)}
 + \frac{1}{2}\D \Psi^{(7/2)}-
   \frac{1}{4}\partial \left[ \Db,\D \right] B^{(2)} \right)
$$
\begin{equation}\label{3}
+  \frac{\partial\left( 5T-2\left[ \Db,\D\right] J
   + B^{(2)} \right)}{\Z} \quad .
\end{equation}
Here  $B^{(2)}(Z),\; \Psi^{(7/2)}(Z),\;{\overline\Psi}^{(7/2)}(Z),\;
U^{(3)}(Z) $ are the composite supercurrents of the spins
$2,\; {7/2},\;{7/2},\;3 $, respectively
\begin{eqnarray}
B^{(2)}(Z) & = & \frac{8}{c} J^2 \nonumber \\
{\overline\Psi}^{(7/2)} & = & \frac{8}{c} \partial\left( J\D J\right)
   -\frac{72}{c}T\D J +\frac{36}{c}\left[\Db ,\D\right]J \D J +
 \frac{8}{c}J\D T -\frac{128}{c^2}J^2\D J
	    +\frac{4}{c}\partial J\D J \nonumber \\
\Psi^{(7/2)} & = & -\frac{8}{c} \partial\left( J\Db J\right)
   -\frac{72}{c} T\Db J +\frac{36}{c}\left[\Db ,\D\right]J \Db J +
\frac{8}{c}J\Db T -\frac{128}{c^2}J^2\Db J
	    -\frac{4}{c}\partial J\Db J \nonumber \\
U^{(3)} & = & \frac{56}{c}J T -\frac{32}{c}J\left[ \Db ,\D \right] J
	 +\frac{128}{c^2}J^3 +\frac{120}{c}\Db J\D J \quad , \label{4}
\end{eqnarray}
where
\begin{equation}\label{5}
\T=\theta_1-\theta_2 \quad , \quad \Tb=\bar\theta_1-\bar\theta_2 \quad ,
 \quad \Z=z_1-z_2+\frac{1}{2}\left( \theta_1\bar\theta_2
-\theta_2\bar\theta_1 \right) \quad ,
\end{equation}
and other definitions have been given in Sec. 3.1.

Next we quote the OPE's of the component currents defined by eq. (3.1).
We use the following notation for the composite currents and their spinor
derivatives
$$
B^{(2)}| = B^{(2)},\; \D B^{(2)}| = \bar{V}^{(5/2)},\;
\Db B^{(2)}| = - V^{(5/2)},\;
\frac{1}{2}\left[ \Db,\D\right] B^{(2)} | = B^{(3)} ,
$$
$$
{\overline\Psi}^{(7/2)}| ={\overline  \Psi}^{(7/2)},\;
\D {\overline \Psi}^{(7/2)}| = {\overline \Lambda}^{(4)},\;
\Db {\overline \Psi}^{(7/2)}| = - \Lambda^{(4)},\;
\frac{1}{2}\left[ \Db,\D\right] {\overline \Psi}^{(7/2)}| =
        {\overline \Psi}^{(9/2)} ,
$$
$$
\Psi^{(7/2)}| = \Psi^{(7/2)},\;
\D \Psi^{(7/2)}| = \Delta^{(4)},\;
\Db \Psi^{(7/2)}| = - {\overline \Delta}^{(4)},\;
\frac{1}{2}\left[ \Db,\D\right] \Psi^{(7/2)}|
 =  \Psi^{(9/2)} ,
$$
$$
U^{(3)}| = U^{(3)},\; \D U^{(3)}| = \overline{\Gamma}^{(7/2)},\;
\Db U^{(3)}| = - \Gamma^{(7/2)},\;
\frac{1}{2}\left[ \Db,\D\right] U^{(3)} | = U^{(4)} \; . \eqno{(A.7)}
$$

Then the OPE's implied for the basic component currents by the
SOPE's (A.1) - (A.3) are as follows
\setcounter{equation}7
\begin{eqnarray}
T(z_1)T(z_2) = \frac{3c}{8\z^4} + \left( \frac{2}{\z^2}+
\frac{\partial }{\z} \right) T  &  \; , \; &
T(z_1)J(z_2) = \left( \frac{1}{\z^2}+\frac{\partial }{\z}\right) J \; ,
 \nonumber\\
T(z_1)G(z_2) = \left( \frac{3}{2\z^2}+\frac{\partial }{\z}\right) G
& \; , \; &
G(z_1)\bar{G}(z_2) = -\frac{c}{4\z^3}+
\left( \frac{1}{\z^2} +
 \frac{\partial }{2\z}\right) J -\frac{ 1}{\z}T  \; ,
 \nonumber\\
G(z_1)G(z_2) = 0
& \; , \; &
J(z_1)G(z_2) = -\frac{1}{\z}G \; ,
 \nonumber\\
J(z_1)J(z_2) = \frac{c}{4\z^2} \; ,
& \;  \; &
\end{eqnarray}
\begin{eqnarray}
T(z_1)\widetilde{T}(z_2) = \left( \frac{2}{\z^2}+
\frac{\partial }{\z}\right) \widetilde{T}
& \; , \;&
T(z_1)W(z_2) = \left( \frac{3}{\z^2}+\frac{\partial }{\z}\right)W \; ,
\nonumber \\
T(z_1)U(z_2) = \left( \frac{5}{2\z^2}+\frac{\partial }{\z}\right) U
& \; , \;&
J(z_1)\widetilde{T}(z_2) = 0 \; ,
\nonumber \\
J(z_1)U(z_2) = -\frac{1}{\z}U
& \; , \;&
J(z_1)W(z_2) = \frac{2}{\z^2}\widetilde{T} \; ,
\nonumber \\
G(z_1)\widetilde{T}(z_2) = \frac{1}{\z}U
& \; , \;&
G(z_1)U(z_2) = 0 \; ,
\nonumber \\
G(z_1)\bar{U}(z_2) = -\frac{1}{\z}W +
\left( \frac{2}{\z^2}+
\frac{\partial}{2\z}\right) \widetilde{T}
& \; , \;&
G(z_1)W(z_2) = \left( \frac{5}{2\z^2}+\frac{\partial }{2\z} \right) U \; ,
\end{eqnarray}
\begin{eqnarray}
\widetilde{T}(z_1)\widetilde{T}(z_2) & = & -\frac{3c}{2\z^4} +
\left( \frac{2}{\z^2}+
\frac{\partial }{\z} \right)
\left( 5\widetilde{T}-4T+B^{(2)}\right) \;, \nonumber \\
U(z_1)\widetilde{T}(z_2)& = & -\left(  \frac{12}{\z^3}+
\frac{8\partial }{\z^2}+\frac{3\partial^2 }{\z}\right) G
+\left( \frac{5}{\z^2}+\frac{3\partial }{\z}\right) U
+\frac{1}{\z}\Psi^{(7/2)}
+\frac{1}{\z^2} V^{(5/2)} \; , \nonumber \\
W(z_1)\widetilde{T}(z_2) & = & -\left(
\frac{12}{\z^4}+\frac{12\partial }{\z^3}
+\frac{6\partial^2 }{\z^2} +\frac{2 \partial^3 }{\z}\right) J
+\left( \frac{3}{\z^2} +\frac{2\partial }{\z}\right) W
+\left( \frac{1}{\z^2} +\frac{\partial }{2\z}\right)U^{(3)} \nonumber \\
&&+\frac{1}{2\z}\left( \Delta^{(4)}- \Lambda^{(4)}-\partial B^{(3)} \right) \;
,
     \nonumber \\
U(z_1)U(z_2) & = & -\frac{1}{\z}{\overline \Delta}^{(4)} \; , \nonumber \\
\overline{U}(z_1)U(z_2) & = &\frac{3c}{\z^5}+
\left( \frac{12}{\z^4}+
\frac{6\partial }{\z^3}+ \frac{2\partial^2 }{\z^2}+
\frac{\partial^3 }{2\z}\right) J
+\left( \frac{20}{\z^3}+\frac{10\partial}{\z^2}+
 \frac{3\partial^2}{\z}\right)
\left(T-\frac{1}{2}\widetilde{T}\right) \nonumber \\
&& -\left( \frac{2}{\z^3}+\frac{\partial }{\z^2} \right) B^{(2)}
+\left( \frac{2}{\z^2}+\frac{\partial }{\z} \right)
\left( W+\frac{1}{2}B^{(3)}-\frac{1}{2}U^{(3)} \right)
-\frac{1}{2\z} \left( \Lambda^{(4)}+\Delta^{(4)}\right) \; , \nonumber \\
W(z_1)U(z_2) &=& - \left( \frac{30}{\z^4}+
\frac{20\partial }{\z^3}+\frac{15\partial^2 }{2\z^2}+
\frac{2\partial^3 }{\z}\right) G+
\left( \frac{5}{\z^3} + \frac{3\partial }{\z^2}+
\frac{\partial^2 }{\z} \right) U \nonumber \\
&&+\left( \frac{1}{\z^3}-\frac{\partial^2 }{4\z}\right)V^{(5/2)}
+\left( \frac{1}{\z^2}+\frac{\partial}{2\z} \right)
\left(\Gamma^{(7/2)} +\frac{1}{2}\Psi^{(7/2)}\right)-
\frac{1}{2\z}\Psi^{(9/2)} \; , \nonumber \\
W(z_1)W(z_2) & =& -\frac{15c}{2\z^6}
+\left( \frac{3}{\z^4}+\frac{3\partial }{2\z^3}
-\frac{\partial^3}{8\z}\right) B^{(2)} -
\left( \frac{15}{\z^4}+\frac{15\partial}{2\z^3}+
\frac{9\partial^2}{4\z^2}+
\frac{\partial^3}{2\z}\right)\left( 4T-\widetilde{T}\right)
\nonumber \\
&& +\left( \frac{1}{2\z^2}+
\frac{\partial}{4\z}\right)
\left(2U^{(4)}+\Lambda^{(4)} +\Delta^{(4)}\right) \;.\label{lastope}
\end{eqnarray}

\setcounter{equation}0
\vspace{0.7cm}

{\Large\bf Appendix B}
\def\theequation{B.\arabic{equation}}

\begin{center}
{\large\bf Basic structure relations of $sW_3^{\infty}$}
\end{center} \vspace{0.3cm}

Here we present the basic (anti)commutation relations of
the superalgebra $sW_3^{\infty}$ in terms of generators.

The generators are defined in a
standard way as Laurent modes of the currents
\begin{equation}
J^{(s)}_n = \frac{1}{2\pi i} \oint dx x^{n+s-1} J^{(s)}(x) \;,
\end{equation}
where $J^{(s)}$ is a current of the spin $s$. Using this
definition, from OPE's (A.8) - (A.10) one obtains the following
(anti)commutation
relations

\begin{eqnarray}
\left[ L_{n} , L_{m}\right] & = &
(n-m)L_{n+m}+\frac{c}{16}\left(n^{3}-n\right)\delta_{n+m,0} \nonumber \\
\left[ L_{n} , G_{r}\right] & = &
\left(\frac{n}{2}-r\right)G_{n+r} \nonumber \\
\left[ L_{n} , J_{m}\right] & = & -mJ_{n+m} \nonumber \\
\left\{G_{r} , \bar{G}_{s}\right\} & = & -L_{r+s}+\frac{r-s}{2}J_{r+s}-
\frac{c}{8}\left(r^{2}-\frac{1}{4}\right)\delta_{r+s,0} \nonumber \\
\left[ J_{n} , G_{r}\right] & = & -G_{n+r} \nonumber \\
\left[ J_{n} , J_{m}\right] & = & \frac{nc}{4}\delta_{n+m,0} \nonumber \\
\left\{G_{r} , G_{s}\right\} & = & \left\{ \bar{G}_{r} ,
\bar{G}_{s}\right\}= 0\;,
\end{eqnarray}

\begin{eqnarray}
\left[L_{n} , \widetilde L_{m}\right] & = &
 (n-m)\widetilde L_{n+m} \nonumber \\
\left[L_{n} , U_{r}\right] & = &
\left( \frac{3n}{2}-r \right)U_{n+r}\nonumber \\
\left[L_{n} , W_{m}\right] & = & (2n-m)W_{n+m} \nonumber \\
\left[G_{r} , \widetilde L_{n}\right] & = & U_{r+n} \nonumber \\
\left\{G_{r} , U_{s}\right\} & = & 0 \nonumber\\
\left\{G_{r} , \bar{U}_{s}\right\} & = &
    -W_{r+s}+\frac{3r-s}{2}\widetilde L_{r+s} \nonumber \\
\left[G_{r} , W_{n}\right] & = &
    \left( 2r-\frac{n}{2}\right) U_{r+n} \nonumber \\
\left[J_{n} , \widetilde L_{m}\right] & = & 0 \nonumber \\
\left[J_{n} , U_{r}\right] & = & - U_{n+r} \nonumber \\
\left[J_{n} , W_{m}\right] & = & 2n\widetilde L_{n+m}\;,
\end{eqnarray}

\begin{eqnarray}
\left[\widetilde L_{n} , \widetilde L_{m}\right] & = &
(n-m)\left[ 5{\widetilde{L}}_{n+m}-4L_{n+m}+B^{(2)}_{n+m}\right ]
-\frac{c}{4}\left(n^{3}-n\right)\delta_{n+m,0} \nonumber \\
\left\{U_{r} , U_{s}\right\} & = &
  -{\overline \Delta}^{(4)}_{r+s} \nonumber \\
\left\{\bar{U}_{r} , U_{s}\right\} & = &
\frac{(r-s)}{4}( 2r^2 + 2s^2 - 5 ) J_{r+s}
+\left( 3r^2+3s^{2}-4rs-\frac{9}{2} \right)
\left( L_{r+s}-\frac{1}{2}{\widetilde L} _{r+s}\right) \nonumber  \\
& &+\frac{(r-s)}{2}\left( 2W+B^{(3)}-U^{(3)} \right)_{r+s}
+\left(r+\frac{3}{2}\right)\left(s+\frac{3}{2}\right)B^{(2)}_{r+s}
-\frac{1}{2}\left(\Lambda^{(4)}+\Delta^{(4)}\right)_{r+s} \nonumber \\
& &+\frac{c}{8}\left(r^{2}-\frac{1}{4}\right)\left(r^{2}-\frac{9}{4}\right)
\delta_{r+s,0}
 \nonumber \\
\left[W_{n} , \widetilde L_{m}\right] & = &
(n-2m)W_{n+m} +2m(m^2-1)J_{n+m}+\frac{n-m+1}{2}U^{(3)}_{n+m} \nonumber \\
& & +\frac{n+m+3}{2}B^{(3)}_{n+m} +\frac{1}{2} \left(\Delta^{(4)}-
 \Lambda^{(4)} \right)_{n+m} \nonumber \\
\left[ U_{r}, \widetilde L_{n} \right] & = &
(2r-3n)U_{r+n}-\left(r^2+3n^2-2rn - \frac{9}{4}\right)G_{n+r}+
\Psi^{(7/2)}_{r+n}+\left( r+\frac{3}{2}\right) V^{(5/2)}_{r+n} \nonumber \\
\left[W_{n} , W_{m}\right] & = &
(n-m)\left(4-n^{2}-m^{2}+\frac{mn}{2}\right)
\left(2L_{n+m}-\frac{1}{2}\widetilde L_{n+m}\right)+
\frac{n-m}{4}\left( 2U^{(4)}+\Lambda^{(4)}+\Delta^{(4)} \right) \nonumber \\
& & -\frac{(n-m)(14+9m+m^2+9n+4mn+n^2)}{8}B^{(2)}_{n+m}
-\frac{c}{16}(n^{3}-n)(n^{2}-4)\delta_{n+m,0} \nonumber \\
\left[W_{n} , U_{r}\right] & = &
\left(r^{2}+\frac{n^{2}}{2}-nr-\frac{5}{4}\right)U_{n+r}-
\left\{\frac{n^{3}}{2}-2r^{3}+\frac{3}{2}nr^{2}-n^{2}r
 -\frac{19}{8}n+\frac{9}{2}r\right\}G_{n+r}
 -\frac{1}{2}\Psi^{(9/2)}_{n+r}\nonumber \\
& &+\frac{4n^2-4r^2-24r-8nr-19}{16}V^{(5/2)}_{n+r}
+\frac{2n-2r+1}{4}\left( \frac{1}{2}\Psi^{(7/2)}+
\Gamma^{(7/2)} \right)_{n+r}\;.
 \label{endcomm}
\end{eqnarray}

Any relation involving higher spin composite generators can be evaluated by
making use of these basic relations and analogous ones involving the
generators $\bar G$ and $\bar U$. These latter relations follow from
those with $G$ and $U$ via the substitutions
\begin{eqnarray}
L & \rightarrow & L  \; , \;
J \; \rightarrow \;-J  \;,\;
W \; \rightarrow \; -W\;,
{\widetilde L}\; \rightarrow \;  {\widetilde L} \nonumber \\
G & \rightarrow & {\bar G} \;,\;
U \; \rightarrow \; -{\bar U}\;.
\end{eqnarray}

\setcounter{equation}0
\vspace{1.3cm}

{\Large\bf Appendix C}
\def\theequation{C.\arabic{equation}}

\begin{center}
{\large\bf The Cartan form coefficients $(l_0)_t$,
$({\tilde l}_0)_t$, $(j_0)_t$ and $(w_0)_t$ }
\end{center}
\vspace{0.3cm}

\begin{eqnarray}
 (l_0)_t && =  - 8\tilde u (2\phi + 3\chi\bar\chi )
               +48 v_2
      +(4\bar\mu +2\phi\bar\chi -6u\bar\chi-12\tilde u \bar\chi+
           6v_1\bar\chi +6\bar\nu ) \xi \nonumber \\
    & & + (4\mu -2\phi\chi -6u\chi-12\tilde u \chi-
           6v_1\chi -6\nu ) \bar\xi
               +24v_1(u+\tilde u )
               +36(\chi\bar\nu +\bar\chi \nu ) \nonumber \\
  & & +\alpha\left( -4\phi_2-2\bar\mu\chi+2\mu\bar\chi+4 u\phi \right)\;,
\end{eqnarray}

\begin{eqnarray}
({\tilde l}_0)_t && =   -4\phi_2 -4\mu\bar\chi +4\bar\mu \chi +
          (2u+10\tilde u )(2\phi +3\chi\bar\chi)
	  -6v_1(u+\tilde u )-18(\chi\bar\nu +\bar\chi \nu) \nonumber \\
      & & - 12v_2 + 3(u+\tilde u )\xi\bar\xi
        -(6\tilde u \chi-3 v_1\chi + 6\nu )\bar\xi -
           (6\tilde u \bar\chi +3 v_1\bar\chi -6\bar\nu )\xi  \nonumber \\
     & & - (2\bar\mu +\phi\bar\chi-3u\bar\chi-6\tilde u \bar\chi+3v_1 \bar\chi+
      3\bar\nu )\xi
     - (2\mu -\phi\chi-3u\chi-6\tilde u \chi-3v_1 \chi-
      3\nu )\bar\xi \nonumber \\
   & & +\alpha\left( 4{\tilde u}\phi -(\mu+\phi\chi)\bar\xi-
      (\bar\mu-\phi\bar\chi )\xi \right)\;,
\end{eqnarray}

\begin{eqnarray}
(j_0)_t && =  12{\tilde u}_2+6(\chi\bar\nu-\bar\chi\nu )
 -\xi\bar\xi (3v_1-2\phi-3\chi\bar\chi ) \nonumber \\
     & & - (2\bar\mu +\phi\bar\chi-3u\bar\chi-6\tilde u \bar\chi+3v_1 \bar\chi+
      3\bar\nu )\xi
     + (2\mu -\phi\chi-3u\chi-6\tilde u \chi-3v_1 \chi-
      3\nu )\bar\xi \nonumber \\
  & & +\alpha\left( 3(\phi^2-u_2)+\mu\bar\chi+\bar\mu\chi+\phi\chi\bar\chi+
        \phi\xi\bar\xi \right)\;,
\end{eqnarray}

\begin{eqnarray}
(w_0)_t && =  -6u_2-6{\tilde u}_2 -4\bar\mu\chi-4\mu\bar\chi+2\phi\chi\bar\chi+
           6(u+\tilde u)^2 -\frac{27}{2}v_1^2 +3v_1(2\phi+3\chi\bar\chi)
             \nonumber \\
     & & +6(\chi\bar\nu -\bar\chi\nu)
  +\frac{3}{2}\xi\bar\xi(v_1-2\phi-3\chi\bar\chi) -
        (6\tilde u \chi-3 v_1\chi + 6\nu )\bar\xi +
           (6\tilde u \bar\chi +3 v_1\bar\chi -6\bar\nu )\xi  \nonumber \\
     & & + (2\bar\mu +\phi\bar\chi-3u\bar\chi-6\tilde u \bar\chi+3v_1 \bar\chi+
      3\bar\nu )\xi
     - (2\mu -\phi\chi-3u\chi-6\tilde u \chi-3v_1 \chi-
      3\nu )\bar\xi \nonumber\\
   & & +\alpha\left( 6v_1\phi -\phi\xi\bar\xi+(\bar\mu-\phi\bar\chi)\xi -
     (\mu+\phi\chi )\bar\xi \right) \;.
\end{eqnarray}
\vspace{0.3cm}

\begin{center}
{\large\bf The expressions for some higher spin coset superfields}
\end{center} \vspace{0.3cm}

\begin{eqnarray}
v_2 & =& \frac{1}{4} \left( \partial v_1+2\tilde u v_1 +2u v_1-\chi\bar\nu-
	\bar\chi\nu +\tilde u \xi\bar\xi +
	\frac{1}{2} ( 2\mu-\phi\chi-\partial\chi-u\chi)\bar\xi
         \right. \nonumber \\
 & & \left.  +
        \frac{1}{2}(2\bar\mu +\phi\bar\chi-\partial\bar\chi-u\bar\chi) \xi
         \right)\;, \\
u_2 & = & \frac{1}{3}( \D\mu+\Db\bar\mu + \phi^2 )\;,  \\
\phi_2 & =&  \frac{1}{2} \partial \phi\;.
\end{eqnarray}

\end{document}